\documentclass[useAMS,usenatbib,aas_macros]{mn2e} 
\usepackage {graphicx,subfigure,times,aas_macros}

\newcommand{\equ}[1]{eq.~(\ref{eq:#1})}

\newcommand{\se}[1]{\S\ref{sec:#1}}

\newcommand{\Fig}[1]{Figure~\ref{fig:#1}}

\newcommand{\tab}[1]{Table~\ref{tab:#1}}
\newcommand{\be}{\begin{equation}}
\newcommand{\ee}{\end{equation}}
\newcommand{\bea}{\begin{eqnarray}}
\newcommand{\eea}{\end{eqnarray}}

\newcommand{\msun}{{\rm M}_\odot}
\newcommand{\Msun}{M_\odot}

\newcommand{\ifm}[1]{\relax\ifmmode#1\else$\mathsurround=0pt #1$\fi}
\newcommand{\kms}{\ifmmode\,{\rm km}\,{\rm s}^{-1}\else km$\,$s$^{-1}$\fi}

\newcommand{\Mpc}{\,{\rm Mpc}}
\newcommand{\kpc}{\,{\rm kpc}}
\newcommand{\pc}{\,{\rm pc}}
\newcommand{\Gyr}{\,{\rm Gyr}}

\newcommand{\Myr}{\,{\rm Myr}}

\newcommand{\ltsima}{$\; \buildrel < \over \sim \;$}
\newcommand{\lsim}{\lower.5ex\hbox{\ltsima}}
\newcommand{\gtsima}{$\; \buildrel > \over \sim \;$}
\newcommand{\gsim}{\lower.5ex\hbox{\gtsima}}

\def\la{\langle}

\def\omm{\Omega_{\rm m}}
\def\oml{\Omega_{\Lambda}}
\def\omb{\Omega_{\rm b}}

\def\cmc{\,{\rm cm}^{-3}}
\def\cms{\,{\rm cm}^{-2}}

\def\Mv{M_{\rm v}}

\def\Ms{M_*}

\newcommand{\VLone}{VL01} 
\newcommand{\VLtwo}{VL02} 
\newcommand{\VLthree}{VL03} 
\newcommand{\VLfour}{VL04} 
\newcommand{\VLfive}{VL05} 
\newcommand{\VLsix}{VL06}
\newcommand{\VLeight}{VL08}
\newcommand{\VLten}{VL10}
\newcommand{\VLeleven}{VL11}
\newcommand{\VLtwelve}{VL12}

\newcommand{\SFGone}{SFG1}
 \newcommand{\SFGtwo}{SFG2}
 \newcommand{\SFGfour}{SFG4}
 \newcommand{\SFGfive}{SFG5}
 \newcommand{\SFGseven}{SFG7}
 \newcommand{\SFGeight}{SFG8}
 \newcommand{\SFGnine}{SFG9}

\def\fj{f_J}
\def\re{r_e}
\def\SigmaOneKpc{\Sigma_{1}}

\usepackage{color}

\begin{document} 

\large 

\title[compact spheroids]
{Early formation of massive, compact, spheroidal galaxies with classical profiles by violent disc instability or mergers}

\author[Ceverino et al.]
{Daniel Ceverino$^1$\thanks{E-mail: daniel.ceverino@uam.es}, 
Avishai Dekel$^2$,
Dylan Tweed$^{2,3}$,
Joel Primack$^4$ \\ 
$^1$Departamento de F{\'i}sica Te{\'o}rica, Universidad Aut{\'o}noma de Madrid, 28049 Madrid, Spain \\
$^2$Racah Institute of Physics, The Hebrew University, Jerusalem 91904, Israel \\
$^3$Center for Astronomy and Astrophysics, Shanghai Jiao Tong University, Shanghai 200240, China \\
$^4$Department of Physics, University of California, Santa Cruz, CA, 95064, USA 
}

\date{}

\pagerange{\pageref{firstpage}--\pageref{lastpage}} \pubyear{0000}

\maketitle

\label{firstpage}

\begin{abstract}

We address the formation of massive stellar spheroids between redshifts $z=4$ and 1 using a suite of AMR hydro-cosmological simulations.
The spheroids form as bulges, and the spheroid mass growth is partly driven by violent disc instability (VDI) and partly by mergers.
A kinematic decomposition to disc and spheroid yields that  
the mass fraction in the spheroid is between 50\% and 90\%  and is roughly constant in time,
consistent with a cosmological steady state of VDI discs that are continuously fed from the cosmic web.
The density profile of the spheroid is typically ``classical", with a Sersic index $n = 4.5\pm 1$, 
independent of whether it grew by mergers or VDI
and independent of the feedback strength.
The disc is characterized by $n=1.5\pm 0.5$, and the whole galaxy by $n=3\pm 1$.
The high-redshift spheroids are compact due to the dissipative inflow of gas and the high universal density.
The stellar surface density within the effective radius of each galaxy as it evolves remains roughly constant in time after its first growth.
For galaxies of a fixed stellar mass,  the surface density is higher at higher redshifts.
\end{abstract}


\section{Introduction}
\label{sec:intro}

%
%
Observations started revealing the properties of massive galaxies during their most active phase of assembly and star formation.  These are galaxies with stellar masses between $\Ms=10^{10} $ and $10^{11} \ \msun$ at redshifts $z=1-4$,
a few Gyrs period during which more than half of the present-day stellar mass was formed  \citep{Dickinson03, Reddy08,Madau14}.

%
%
A significant fraction of the massive star-forming galaxies    
can be characterized as thick, turbulent, extended, rotating discs  
that are highly perturbed by transient elongated features and giant clumps 
\citep{Genzel06,Elmegreen06,Genzel08,Stark08,Law09,Forster09,Forster11a,Wuyts12,Guo12,Guo14}.  
First dubbed ``clump-cluster" and ``chain" galaxies, based on their   
face-on or edge-on rest-frame UV images  
\citep{Cowie95,Bergh96,Elmegreen04b,Elmegreen05,Elmegreen07}, 
a significant fraction of these galaxies ($>50\%$ for the more massive
ones) are confirmed by spectroscopic measurements to be rotating and extended  
discs \citep{Genzel06,Shapiro08,Forster09}. 

%
%
A substantial fraction
of these high-$z$ clumpy discs also host a central, massive stellar bulge
\citep{Genzel08,Elmegreen09,Szomoru11, Szomoru12,Genzel14}.  
This compact bulge may passively evolve into a compact spheroid, as the outer star-forming regions fade away \citep{Kriek09b}.
Therefore, observations indicate an evolutionary link between clumpy discs and compact spheroids at high redshifts.

%
High-redshift spheroids are much more compact than galaxies of similar mass at z=0 \citep{Daddi05, 
Trujillo06,Toft07, vDokkum08,Cimatti08}.
These galaxies with stellar masses of $\sim$$10^{11} \ \msun$ at $z=1.5-2$ have typical effective radii of $\sim$2 kpc, a factor $\sim$3 smaller than present-day spheroids
and a factor $\sim$2 lower than extended discs with similar mass at the same redshift \citep{Patel13}.
Therefore, the spheroids are significantly denser than other high-z galaxies.
Moreover, they show steep surface brightness profiles \citep{Cassata10, Conselice11, Szomoru12}, consistent with the ``classical'', de-Vaucouleurs profile \citep{Vaucouleurs}.
%
%
``Blue nuggets'' \citep{Barro13, Williams14, Barro14, Barro14b} are examples of compact and dense, star-forming galaxies at high redshifts, $z>1.5$.
Observations indicate that these galaxies are continuously being formed at $z = 2-3$,  by fast and highly dissipative formation processes, which generate the observed  high stellar surface densities and steep mass profiles.

%
%

%
%
%
%

%
%

 
While major mergers do lead to spheroids
\citep{Robertson06,Cox06,DekelCox,Hopkins06,Naab07}
and compact galaxies \citep{Wuyts10},
the major-merger rate is not enough for producing spheroids
in the observed abundance 
\citep[][and references therein]{dekel09,Genel08,Stewart09},
so an additional and complementary mechanism 
must have been working efficiently at high redshifts.
Here, we propose violent disc instability (VDI), associated with intense gas in-streaming including wet minor mergers, as another viable mechanism for producing compact spheroids at high redshifts.

The gravitational fragmentation of gas-rich and turbulent galactic discs  
into giant clumps has been addressed by simulations in the idealized   
context of an isolated galaxy  
\citep{Noguchi99, Immeli04, Elmegreen05, Bournaud07, Elmegreen08a, Bournaud09}, 
and in a cosmological context, using analytic theory 
\citep[][DSC]{DSC} and cosmological simulations  
\citep{Agertz09b,CDB,Genel12,Ceverino12,Mandelker,Moody}.  
According to the standard Toomre instability analysis \citep{toomre64}, 
a rotating disc becomes unstable to local gravitational  
collapse if its surface density is sufficiently high for its self-gravity to 
overcome the forces induced by rotation and velocity dispersion that resist  
the collapse.  
If the disc is maintained in a marginally unstable state with the maximum 
allowed velocity dispersion,  
and the cold fraction of the disc is about one third of the total mass 
within the disc radius,  
the perturbed, clumpy disc induces mass inflow  
into the disc centre \citep[DSC;][]{Dekel13}.

This mass inflow is partly due to clump migration \citep{Noguchi99,Elmegreen05, Bournaud07},
and is partly a smooth inflow of disc gas outside the clumps \citep{Gammie01,Bournaud11,Forbes12},
generated by torques within the perturbed disc, that drive angular momentum outward.
One should emphasize that the inflow in the disc is a robust feature of the instability, not limited to clump migration.
In the high-$z$ gas-rich discs,  this gas inflow is fast, as it operates on 
the disc dynamical timescale.
During this timescale, turbulence dissipates, but the dissipation is compensated by the energy gained by the inflow down the potential well. 
The gravitational energy drives turbulence
that helps maintaining the disc in the marginally unstable state 
\citep{krumholz_burkert10,Cacciato12,Forbes12,Forbes14}.  
The mass inflow contributes to the growth of a central bulge, which in turn 
tends to stabilize the disc,
and it keeps a roughly constant spheroid mass fractions of $S/T \sim 0.6-0.7$ \citep{Dekel13}.
The continuous intense streaming of fresh  
cold gas from the cosmic web into the disc  
\citep{bd03,keres05,db06,ocvirk08,dekel09} 
maintains the high surface density 
and keeps the instability going for several Gyr in a quasi-stationary state (DSC). 
The analytical estimates of the gas inflow within the disc imply that the rate of gas draining from the disc to the central spheroid is expected to be comparable to the rate of replenishment by freshly accreting gas \citep{Dekel13}.
This inflow continues as long as the gravitational disc instability is ongoing.
Therefore, VDI is an efficient mechanism to build  a compact and dense spheroid at high redshifts.

The main goal of this paper is to characterize the morphological properties of the spheroids produced by VDI and by mergers in cosmological simulations.
Isolated-disc simulations have yielded results of mixed nature. 
Sticky-particles simulations \citep{Elmegreen08a} reported the formation of VDI-driven bulges with a classical, de-Vaucouleurs mass profile.
In contrast, \citet{inoue12} reported SPH simulations that produced pseudo-bulges with nearly exponential profiles. Although the different numerical techniques may explain this discrepancy, the selected initial conditions of the idealized examples are likely to play an important role in the subsequent evolution of the isolated disc into a central bulge. 
Therefore, the study of the mass profiles of the spheroids produced in cosmological simulations with primordial initial conditions is crucial for distinguishing  between these two results.
Another goal of this paper is the characterization of the mass and size growth of massive galaxies at high redshift.
Making detailed predictions of the early growth of todays massive galaxies
will be very useful for the interpretation of  observed galaxy properties from ongoing large surveys at high redshift.

This paper is organized as follows. 
\se{models} describes the simulations used in this paper.
\se{example} shows an example of the formation of a massive and compact spheroid.
\se{sample} gives an overview of the spheroid growth due to VDI and mergers.
\se{morpho} characterizes the mass profiles of the spheroid and disc components.
\se{MassSize} discusses the mass and size growth of the simulated galaxies.
\se{fixedmass} describes the redshift evolution of a sample of galaxies at fixed stellar mass.
Finally, \se{Summary} summarizes the results and discusses them.

\section{Simulations}
\label{sec:models}

We use zoom-in hydro cosmological simulations of a relatively large set of 17 massive 
halos ($\Mv \geq 2 \times 10^{12} \ \msun$ at $z=1$) with an AMR maximum resolution of $70\pc$ or better, evolved till 
$z \sim 1$ in most of the cases. 
This sample was drawn from a bigger set, first used in \citet{Dekel13}.
In addition, we use a couple of simulations with twice better resolution ($35\pc$) and a stronger feedback model, in order to test the effects of resolution and feedback. These two halos were drawn from a bigger set, called the Vela set, first used in \citet{Ceverino13} and Zolotov et al. (in prep). We selected the runs with halo masses at  $z \sim 1$ similar to the halo masses of the main sample of this paper. 

All simulations utilize the  \textsc{ART} code
\citep{Kravtsov97,Kravtsov03}, which accurately follows the evolution of a
gravitating N-body system and the Eulerian gas dynamics using an adaptive mesh.
Beyond gravity and hydrodynamics, the code incorporates at the subgrid level
many of the physical processes relevant for galaxy formation.
They include gas cooling by atomic hydrogen and helium, metal and molecular
hydrogen cooling, photoionization heating by a UV background with partial
self-shielding, star formation, stellar mass loss, metal enrichment of the ISM,
and feedback from stellar winds and supernovae (\se{code}), and radiative feedback in the case of the Vela runs (\se{vela}).

\subsection{Growth of selected halos}
\label{sec:set}

The 17 dark-matter haloes were drawn from N-body simulations
of the $\Lambda$CDM cosmology with the WMAP5 parameters (\se{code}),
in a comoving cosmological box. 
The haloes were selected to have a virial mass in a desired mass   
range at $z=1$ (\tab{1}). 
13 out of 17 halos have very similar virial masses of $2-3 \times 10^{12} \ \msun$, representative of massive but common halos at $z\simeq1$. We will refer to this homogeneous subset of simulations as the main sample throughout the paper. 
The rest of simulations have significantly higher virial masses, so they are less common halos.
\SFGeight \ and \SFGnine \ have a $\Mv=5-6 \times 10^{12} \ \msun$. Finally, \SFGtwo \ and \SFGseven \ are two examples of rare, group-size halos with $\Mv\simeq 10^{13}  \ \msun$ at $z=1$.
Due to the high computational cost, these two runs only reached $z\simeq4$. However, they assembled a halo of $10^{12} \ \msun$ virial mass at this redshift and therefore they can be compared with the runs that reached similar halo mass but at lower redshifts.
Finally, the only other selection criterion was that the selected halos show no ongoing major merger at the selected time. This eliminates less than $10\%$ of the halos.

\begin{table} 
\caption{Set of zoom-in simulations. Columns show the name of the run, the Universe expansion factor (a$=(1+z)^{-1}$) of the last snapshot, the virial and stellar mass at that snapshot, and  the number of dark matter particles in the run.}
 \begin{center} 
 \begin{tabular}{cccccc} \hline 
\multicolumn{2}{c} {Run } \ \ a$_{\rm last}$ & virial mass & stellar mass & \# DM particles  \\
\hline 
\VLone   & 0.37 & $1.9 \times 10^{12}$  & $1.5 \times 10^{11}$  & $11 \times 10^6$ \\ 
\VLtwo    & 0.50 & $2.1 \times 10^{12}$  & $1.3 \times 10^{11}$ &  $ 9.0 \times 10^6$\\
\VLthree  & 0.33 & $1.2 \times 10^{12}$  &$0.7 \times 10^{11}$ &  $12 \times 10^6$ \\  
\VLfour    & 0.42 & $1.2 \times 10^{12}$  &$1.4 \times 10^{11}$ &  $9.2 \times 10^6$ \\ 
\VLfive    & 0.41 & $2.1 \times 10^{12}$  & $1.8 \times 10^{11}$ &  $8.5 \times 10^6$\\ 
\VLsix     & 0.50 & $2.0 \times 10^{12}$  & $1.0 \times 10^{11}$ &  $9.2 \times 10^6$ \\ 
\VLeight   & 0.46 & $1.2 \times 10^{12}$ &$1.6 \times 10^{11}$ &  $14 \times 10^6$ \\ 
\VLten       &  0.50 & $2.8 \times 10^{12}$&$3.5 \times 10^{11}$ & $21 \times 10^6$ \\ 
\VLeleven  & 0.50 & $2.2 \times 10^{12}$&$2.5 \times 10^{11}$ &$13 \times 10^6$ \\ 
\VLtwelve  & 0.50 & $2.1 \times 10^{12}$ &$1.4 \times 10^{11}$ &$15 \times 10^6$  \\ 
\SFGone    & 0.38 & $2.0 \times 10^{12}$ &$2.2 \times 10^{11}$ &$12 \times 10^6$\\
\SFGfour    & 0.38 & $1.3 \times 10^{12}$ &$1.1 \times 10^{11}$ &$12 \times 10^6$   \\  
\SFGfive     & 0.40 & $1.7 \times 10^{12}$ &$1.4 \times 10^{11}$ &$12 \times 10^6$ \\ 
\hline
\SFGnine    & 0.48 &  $ 5.3 \times 10^{12}$  & $5.6 \times 10^{11}$ & $23 \times 10^6$  \\ 
\SFGeight   & 0.35 &  $ 1.4 \times 10^{12}$  & $1.5 \times 10^{11}$ & $30 \times 10^6$  \\ 
\SFGseven & 0.22 &  $ 1.3 \times 10^{12}$  & $1.2 \times 10^{11}$ &  $56 \times 10^6$ \\ 
\SFGtwo      & 0.21 & $  2.1 \times 10^{12}$  & $1.9 \times 10^{11}$ & $53 \times 10^6$ \\ 
\hline
Vela07         & 0.50 & $  1.5 \times 10^{12}$  & $0.7 \times 10^{11}$ & $43 \times 10^6$ \\
Vela33         & 0.39 & $  1.5 \times 10^{12}$  & $0.9 \times 10^{11}$ & $29 \times 10^6$ \\
\hline 
 \end{tabular} 
 \end{center} 
\label{tab:1} 
 \end{table} 
 
 \begin{figure} 
\includegraphics[width =0.47 \textwidth]{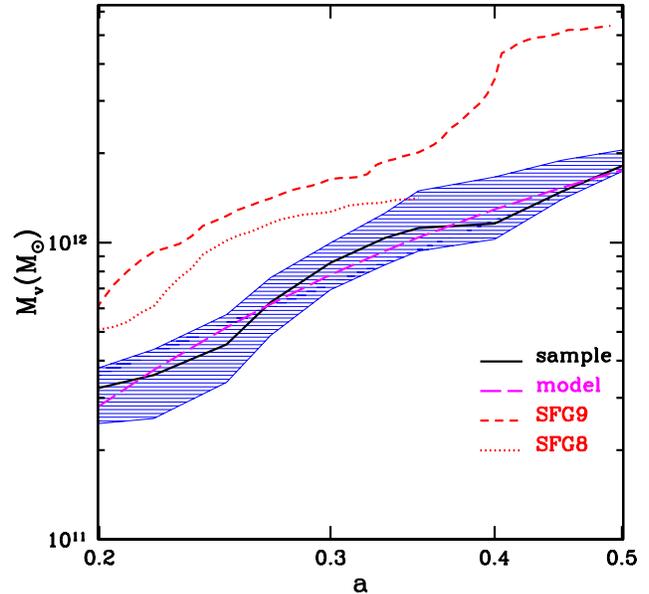}
\caption{Evolution of the virial mass as a function of the Universe expansion factor, a. The black, solid line corresponds to the median of the sample (\VLone \ to \SFGfive) and the blue shade region encompasses the 25\% and 75\% percentiles. Other lines represent more massive runs (\SFGeight \ and \SFGnine). 
The most massive runs, \SFGtwo \ and \SFGseven \ are not shown.
The magenta, long-dashed line shows the prediction from the toy model, \equ{Mh}. }
\label{fig:halos}
\end{figure}
 
 %
 %
 
 The growth of these halos can be explained analytically by a simple toy model, valid for massive halos, $\Mv\simeq10^{12} \ \msun$ at $z\geq1$. The mean specific accretion rate can be approximated in the Einstein-deSitter regime by the following relation, ignoring a weak mass dependency \citep{Dekel13}:
\be
\frac{\dot{M}}{M}=s (1+z)^{5/2}
\label{eq:Mdot}
\ee
A simple integration gives the average growth of the virial mass of a halo as a function of redshift, up to z=1:
\be
M(z)=M(z=1) e^{\alpha (z-1)} \\   , \\  \alpha = - \omm^{-1/2} H_0^{-1}  s
\label{eq:Mh}
\ee
where $H_0=70 \ {\rm Mpc^{-1}} \ {\rm km} \ s^{-1}$ is the Hubble constant and $s =0.030 \Gyr^{-1}$ gives the best fit to halos of $\Mv=10^{12} \ \msun$ from large cosmological simulations \citep{Neistein08, Dekel13}

\Fig{halos} shows the evolution of the virial mass as a function of the Universe expansion factor, $a=1/(1+z)$.
The median of the sample grows significantly from $\Mv\simeq 3 \times 10^{11} \ \msun$ at $z=4$ to $\Mv \simeq 2 \times 10^{12} \ \msun$ at $z=1$.
This growth is well reproduced by the toy model just described, \equ{Mh}, where $M(z=1)=1.7 \times 10^{12} \ \msun$ and $\alpha=-0.61$ give the best fit. The corresponding  $s=0.023 \Gyr^{-1}$ is remarkable similar to the value predicted by the toy model. This is not completely unexpected because the toy model was calibrated using dark matter halos of similar mass, drawn from large cosmological simulations.

 %
 %
 
 The sample is consistent with a roughly constant stellar-to-virial ratio, $\Ms/\Mv \simeq 0.09$, roughly half of the Universal baryonic fraction.
 This is too high by a factor of 2-3 in comparison to what is predicted using abundance matching techniques \citep{Moster12, Behroozi12,Kravtsov14}.
It is known that these simulations produce too many stars at high $z$ due to a lack of strong feedback and perhaps a too-high star-formation efficiency.
This produces an excess of star formation in small halos at high redshifts. It translates into low gas fractions and low star formation rates at $z=2$ by a factor $\sim$2 (CDB). This has implications in the formation of compact spheroids by wet processes (VDI and mergers). These simulations probably underestimate the importance of VDI in the formation of spheroids due to the relatively low gas fractions in the disc, as discussed in CDB. 

The majority of the morphological properties of these massive galaxies are robust to changes in the stellar mass fraction by a factor of a few, especially at $z=1-2$.
 Simulations of halos of similar mass with stronger feedback (Vela07 and Vela33) significantly reduce the stellar fractions ($\Ms/\Mv \simeq 0.04-0.06$) but the morphological properties are qualitatively the same, especially for the progenitors of massive galaxies with $\sim$$10^{11} \Msun$ in stars at $z\simeq1$.
 
\subsection{  \textsc{ART} }
\label{sec:code}

 The  \textsc{ART} code 
\citep{Kravtsov97,Kravtsov03} follows the evolution of a 
gravitating N-body system and the Eulerian gas dynamics using an adaptive mesh 
refinement approach. Beyond gravity and hydrodynamics, the code incorporates 
many of the physical processes relevant for galaxy formation, as described in 
\citet{Ceverino09} and CDB.  These processes, representing subgrid 
physics, include gas cooling by atomic hydrogen and helium, metal and molecular 
hydrogen cooling, and photoionization heating by an UV background with partial 
self-shielding. 
Cooling and heating rates are tabulated for a given gas 
density, temperature, metallicity and UV background based on the  \textsc{cloudy} code 
\citep{Ferland98}, assuming a slab of thickness 1 kpc. An uniform UV background 
based on the redshift-dependent \citet{HaardtMadau96} model is assumed, 
except at gas densities higher than $0.1\cmc$, where a substantially 
suppressed UV background is used
($5.9\times 10^{26}{\rm erg}{\rm s}^{-1}{\rm cm}^{-2}{\rm Hz}^{-1}$) 
in order to mimic the partial self-shielding of dense gas. 
This allows the dense gas to cool down to temperatures of $\sim$300 K. 
The assumed equation of state is that of an ideal mono-atomic gas. 
Artificial fragmentation on the cell size is prevented by introducing 
a pressure floor, which ensures that the Jeans scale is resolved by at least 
7 cells (see CDB). 
 
Star formation is assumed to occur at densities above a threshold of $1\cmc$ 
and at temperatures below $10^4$ K. More than 90\% of the stars form at 
temperatures well below $10^3$ K, and more than half the stars form at 300~K 
in cells where the gas density is higher than $10\cmc$. 
The code implements a stochastic star-formation model that yields a 
star-formation efficiency per free-fall time of 5\%. At the given resolution, 
this efficiency roughly mimics the empirical Kennicutt law \citep{Kennicutt98}. 
The code incorporates a thermal stellar feedback model, in which the combined 
energy from stellar winds and supernova explosions is released as a constant 
heating rate over $40\Myr$ following star formation, the typical age of the 
lightest star that explodes as a type-II supernova. 
The heating rate due to feedback may or may not overcome the cooling 
rate, depending on the gas conditions in the star-forming regions 
\citep{DekelSilk86,Ceverino09}. No shutdown of cooling is implemented. 
We also include the effect of runaway stars by assigning a velocity kick of 
$\sim 10 \kms$ to 30\% of the newly formed stellar particles. 
The code also includes the later effects of type-Ia supernova and 
stellar mass loss, and it follows the metal enrichment of the ISM. 
 
The initial conditions corresponding to each of the selected haloes 
were filled with gas and refined to a much higher resolution on an adaptive 
mesh within a zoom-in Lagrangian volume that encompasses the mass within 
twice the virial radius at $z =1$, roughly a sphere of comoving radius $\sim$1 Mpc. 
This was embedded in a comoving cosmological box of side 57 comoving Mpc for \VLone - \VLsix \ and \SFGone, \SFGfour, and \SFGfive, or 114 comoving Mpc for the runs  \VLeight-\VLtwelve , \SFGtwo , and \SFGseven.
A standard $\Lambda$CDM cosmology has been assumed, with the 
WMAP5 cosmological parameters 
$\omm=0.27$, $\oml=0.73$, $\omb= 0.045$, $h=0.7$ and $\sigma_8=0.82$ 
\citep{WMAP5}. 

The zoom-in regions have been simulated with 
$(8-46)\times 10^6$ dark-matter particles with a minimum mass of 
$6.6 \times 10^5 \ \msun$, while the particles representing stars 
have a minimum mass of $10^4 \ \msun$. 
Each galaxy has been evolved forward in time with the full hydro  \textsc{ART} and 
subgrid physics on an adaptive comoving mesh refined in the dense regions 
to cells of minimum size between 35 and 70 pc in physical units at all times. 
Each AMR cell is refined to 8 cells once it contains a mass in stars and 
dark matter higher than $2\times 10^6 \ \msun$, equivalent to 3 dark-matter 
particles, or it contains a gas mass higher than $1.5\times 10^6 \ \msun$. This 
quasi-Lagrangian strategy ends at the highest level of refinement 
that marks the minimum cell size at each redshift. 
In particular, the minimum cell size is set to 35 pc in physical units 
at expansion factor $a=0.16$ ($z=5.25$), but due to the expansion of the 
whole mesh while the refinement level remains fixed, 
the minimum cell size grows in physical units and becomes 70 pc by 
$a=0.32$ ($z=2.125$). 
At this time we add a new level to the comoving mesh, so the minimum cell 
size becomes 35 pc again, and so on. 
This maximum resolution is valid in 
particular throughout the cold discs and dense clumps, allowing cooling to 
$\sim 300$ K and gas densities of $\sim 10^3\cmc$. 

\subsection{Vela runs }
\label{sec:vela}

The runs Vela07 and Vela33 use radiative feedback in addition to the thermal feedback described above. The model of radiative feedback is described in \citet{Ceverino13}. 
In short, this model adds a non-thermal pressure, radiation pressure, to the total gas pressure in regions where ionizing photons from massive stars are produced and trapped. 
In the current implementation, named RadPre in \citet{Ceverino13}, radiation pressure is included in the cells (and their closest neighbors) that contain stellar particles younger than 5 Myr and whose gas column density exceeds $10^{21}\ \cms.$ 

In addition to feedback, there are two other main differences with respect to the models used in the main sample.
Firstly, the Vela runs use a Chabrier IMF (Chabrier 2005), as opposed to the original Miller-Scalo IMF. 
Secondly, the star formation efficiency is lower by a factor of 3 with respect to the original CDB value. 
At the resolution of these new simulations, this value roughly yields the empirical Kennicutt-Schmidt law \citep{Ceverino13}. 

The initial conditions of these Vela runs contain $30-40 \times 10^6$ dark matter particles with a minimum mass of 
$8.3 \times 10^4 \ \msun$, while the particles representing stars have a minimum mass of $10^3 \ \msun$. 
The maximum spatial resolution is between 17-35 proper pc. More details can be found in \citet{Ceverino13} and Zolotov et al. (in prep).

\begin{figure*} 
\includegraphics[width =0.9\textwidth]{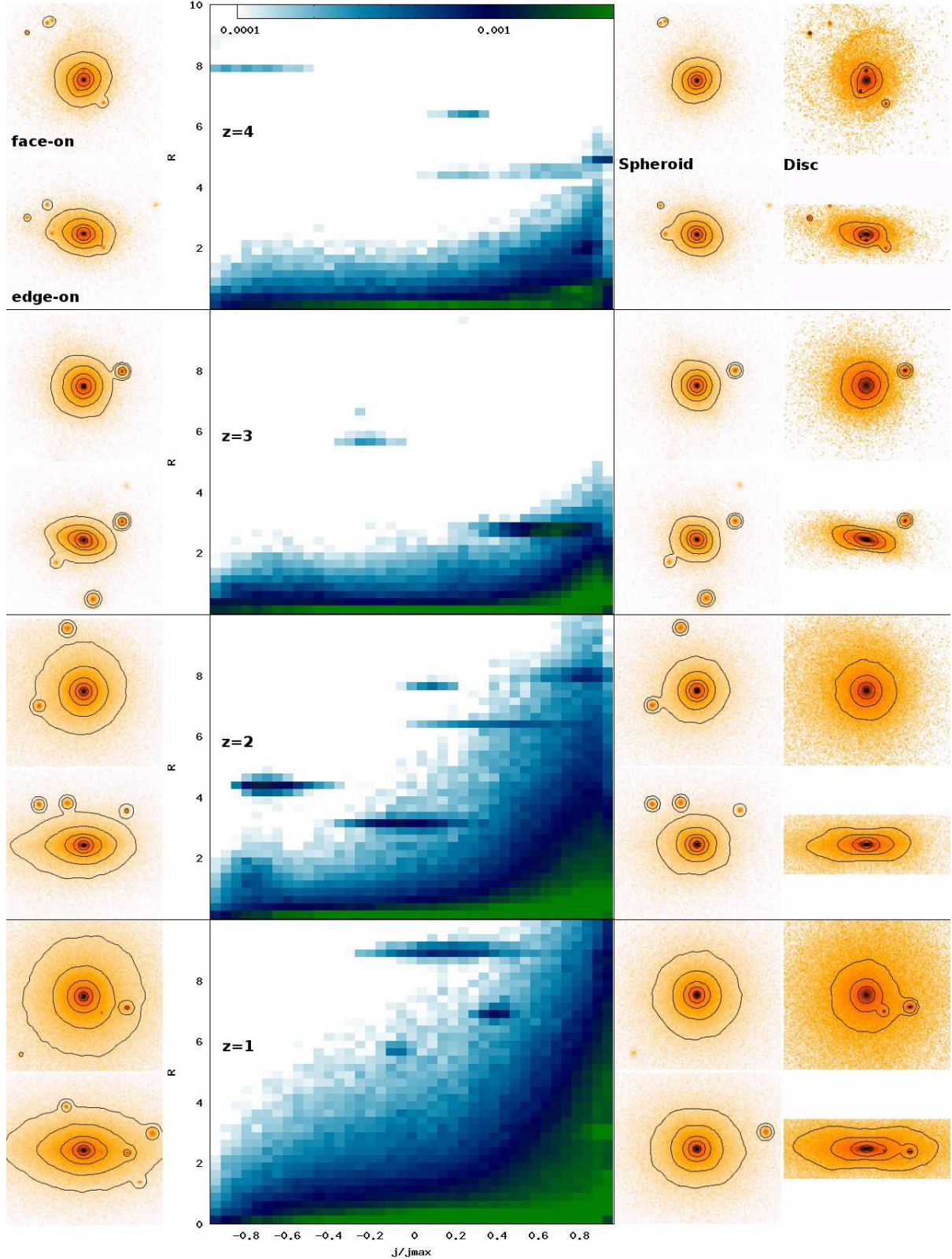}
\caption{Evolution of galaxy \VLsix \ at different redshifts: $z=$4,3,2,1. Left:  face-on (top) and edge-on (bottom) views of stellar surface density. 
The maximum value of the color palette always corresponds to the maximum value of the stellar surface density (log scale) in each individual image.
On the other hand, the isodensity contours are always the same and they are spaced 0.5 dex, starting from log$(\Sigma / \msun \kpc^{-2})=8.5$. 
The size of the panels is always $ 10 \times 10 \kpc$.
Middle: mass-weighted distribution in the plane of $\fj=j_z/j_{\rm max}$ and face-on radius R.
The color code in each bin corresponds to the stellar mass in that bin divided by the total stellar mass at $z=1$.
Right: surface density maps of the spheroid, $\fj<0.7$ (left) and disc, $\fj>0.7$ and $|z|<2 \kpc$ (right) components in the face-on (top) and edge-on (bottom) views.}
\label{fig:VL06}
\end{figure*}

\section{ A typical massive compact spheroid}
\label{sec:example}

In this section we focus on one of our simulated galaxies and use it to illustrate our method of analysis.  In \se{VL06} we visualize the formation of a compact spheroid, in \se{BD} we describe the decomposition into a spheroid and a disc, and in \se{profiles} we address the face-on stellar surface-density profile.

\subsection{Evolution of a single galaxy}
\label{sec:VL06}

We choose galaxy \VLsix \ as a typical example of a massive ($\Ms \sim 10^{11} \  \msun$ at $z=1$) and compact elliptical galaxy with an effective radius of 2.7 \kpc  \ at $z=1$.
The left column of \Fig{VL06} shows the overall morphology of \VLsix \ at different redshifts.
At $z=4$, 
the stellar system appears to be a rather flattened oblate small system.
This indicates that the mass in a rotating component may be non-negligible. 
The gas component shows a thick and turbulent disc at all times, as shown in CDB, \citet{Ceverino12} and \citet{Mandelker}.
The galaxy grows at lower redshifts and it evolves into a larger elliptical galaxy with an overall flattened shape.
The final snapshot at $z=1$ resembles a compact counterpart of ordinary flattened elliptical galaxies in the local Universe. It has an ellipticity of 0.46, as measured at the isodensity contour of $10^9 \ \msun \kpc^{-2}$, at roughly the galaxy effective radius.

From the distribution of specific angular momentum, it is apparent that at any time the galaxy can be decomposed into two  distinct components: a `disc' component with a relatively high level of rotation support plus a spheroidal component with little rotation support.
It is important to disentangle these two components because they have  different kinematics and morphologies so they must  have formed by different processes.
If we want to focus on the formation mechanism of the spheroidal component, we need first to disentangle it  from the disc in order to study the clues of its formation, imprinted in its intrinsic morphology.

\subsection{Spheroid/Disc decomposition using stellar kinematics}
\label{sec:BD}

We follow the procedure discussed in CDB to separate the distribution of stellar particles into spheroid and disc components, using stellar kinematics. 
Per each stellar particle within a face-on projected radius of 10 \kpc \ from the galaxy center, the specific angular momentum parallel to the galaxy angular momentum, $j_z$,  is  compared with the maximum specific angular momentum that a particle with the same velocity and position could have: $j_{\rm max}=|v| \ r$, where $|v|$ is the magnitude of the particle velocity and r is its distance from the center.
This ratio is $\fj=1$ for a perfect circular orbit in the galaxy plane; $\fj=-1$ for a circular retrograde orbit and $\fj=0$ for a radial orbit with zero angular momentum.

The middle column of \Fig{VL06} shows the stellar mass distribution in the plane of $\fj$ and projected radius R at different redshifts (See \citet{Scannapieco09}).
In this plane, a spherical component with total angular momentum equal to zero will show a symmetric distribution around $\fj=0$. On the other hand, a disc composed of prograde circular orbits will show a narrow peak at $\fj=1$. 
In all redshifts, it is possible to see this peak around $\fj=1$, an indication of a disc.
The disc grows in size with time.
On the other hand, the spheroid, centered at $\fj=0$ looks more concentrated than the disc and it extends almost symmetrically  from prograde to retrograde orbits.

The phase-space distributions also show stellar clumps at different radii.
These are the ex-situ and in-situ clumps discussed in \citet{Ceverino12} and \citet{Mandelker}.
The ex-situ clumps of satellite galaxies have  diverse orbits that sometimes differ from circular orbits.
The in-situ clumps are formed by disc instabilities. They orbit well inside the disc in close-to-circular orbits with $\fj\sim1$.
Using only kinematics, it is difficult to distinguish the in-situ clumps from the satellites, because half of these ex-situ clumps also share the disc rotation \citep{Mandelker}.

Based on these distributions and the edge-on views of \Fig{VL06}, we can define the disc and the spheroid components as follows. The spheroid is composed of all stars with $\fj<0.7$.
The disc is composed of all stars with  $\fj>0.7$ and within $2 \kpc$ from the plane of the galaxy, defined by the galaxy total angular momentum within 3 \kpc \ from the galaxy center.
This definition of disc and spheroid, based on a threshold value of $\fj$ is somewhat arbitrary.
The phase-space plots show a continuous distribution of material with low and high $\fj$ values, with no clear sharp boundary.
We tried different values of the $\fj$ threshold, between 0.6 and 0.8, and the main morphological properties of the disc and the spheroid remained similar. 
A higher threshold selects disc material closer to the galaxy plane, making the disc thinner and less massive. The spheroid on the other hand becomes slightly more elongated, but its face-on surface density profile remains the same. This is due to the fact that the bulk of the spheroid mass has low $\fj$ values.

\begin{figure} 
\includegraphics[width =0.49 \textwidth]{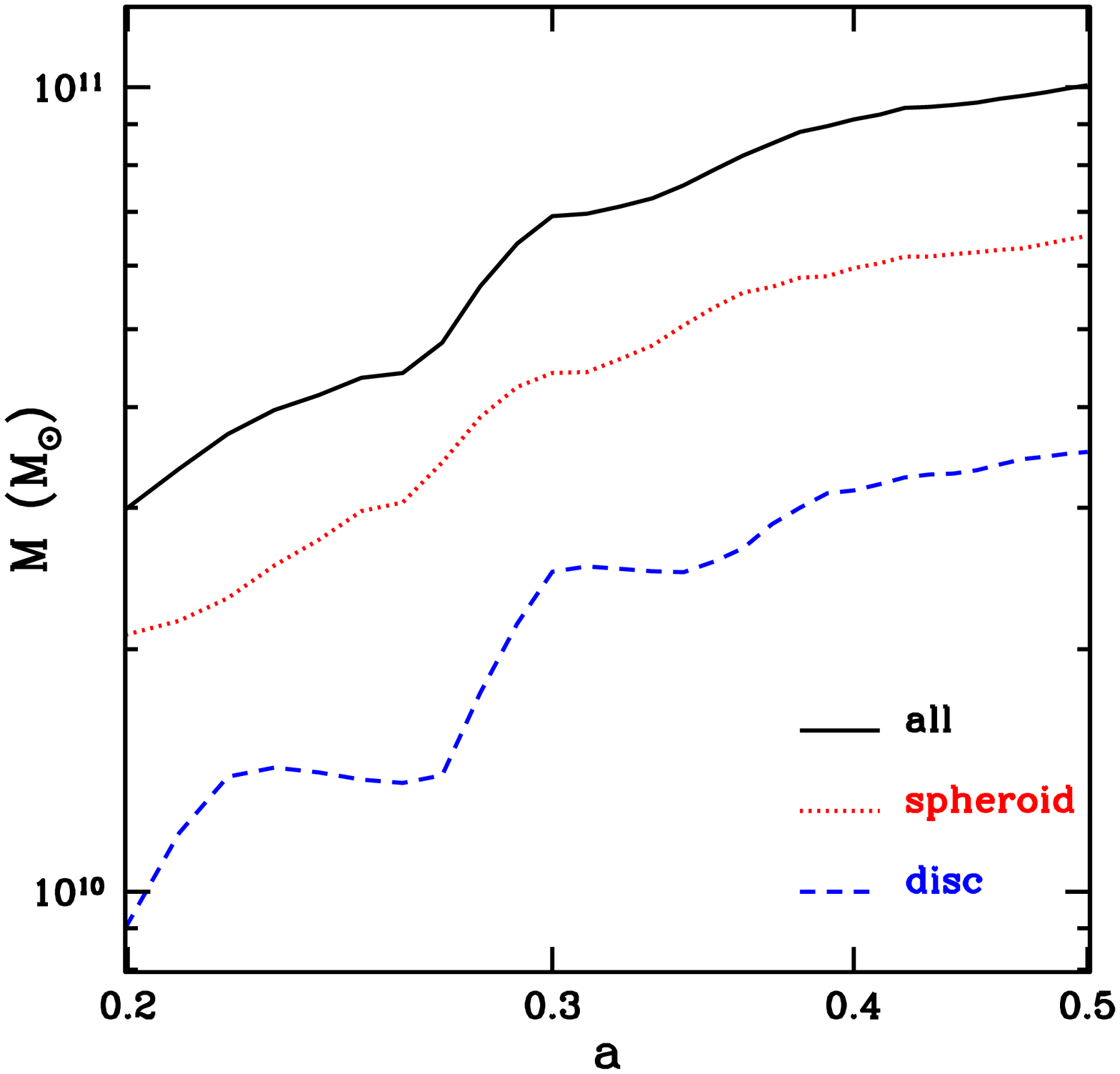}
\caption{Stellar mass growth in the disc, the spheroid, and the whole galaxy, for galaxy \VLsix. The two components grow in a similar rate. Except for a steep rise due a major merger at $a \sim 0.3$, the growth is rather smooth, driven by VDI.}
\label{fig:MsfDVL06}
\end{figure}

The right column of  \Fig{VL06} shows the face-on and edge-on maps of these two components, spheroid and disc, separately.
The spheroid, or the material with relatively low angular momentum in the direction of the galaxy rotation axis, always exhibits a spheroidal shape with little difference between the two orthogonal views.
On the other hand, the disc component, or the material with relatively high angular momentum, always show a flattened shape, although the disc often looks irregular or asymmetric, as well as thick, especially at later times.
These high-z thick stellar discs turn out to be different from the thin discs at low redshift.

The face-on views of the spheroid and the disc also show important differences between these two components. 
Low surface density material outside the last isodensity contour are more prominent in the disc than in the spheroid. 
In other words, the mass fraction in the outskirts of the disc is higher than in the spheroid. This means that the spheroidal component is more concentrated that the disc at all times. This is related to the different formation mechanisms of the disc and the spheroid. The later is formed through violent, gas-rich processes, like VDI and wet mergers that dissipate a considerable amount of angular momentum and orbital energy. 
The stellar disc is formed from high-spin material if the SFR is faster than the inflow rate within the disc \citep{DekelBurkert}.

Both spheroid and disc grow in mass and size with time.
\Fig{MsfDVL06} shows the evolution of the total stellar mass, the mass in the spheroid and in the disc.
The total stellar mass grows from $3 \times 10^{10} \Msun$ at $z=4$ to  $10^{11} \Msun$ at $z=1$.
This is a significant growth by a factor $\sim3$ in $\sim4$ Gyr of evolution.
The relative mass fraction in the two components stays roughly constant: the spheroid is more massive than the disc at all times.
During most of the time, the spheroid grows continuously, rather than in discrete jumps, 
while there is a significant jump in the stellar mass due to a major merger right before $a\simeq0.3$.
A continuous, quasi-steady-state process should dominate the spheroid build-up at these early times.
VDI has been proposed as such mechanism.
If the disc is marginally unstable, there is continuous transfer of material from the disc to the spheroid. This mass transfer is a steady process that can last a few Gyrs, as long as the disc  self-regulates itself in marginal instability. 
During this time, the disc mass that is drained into the central bulge is replenished by continuous accretion of fresh gas from the cosmic web, as described analytically \citep[DSC;][]{DekelBurkert} and shown in simulations \citep[CDB;][Zolotov et al.]{Dekel13}.

\begin{figure*} 
\includegraphics[width =0.95 \textwidth]{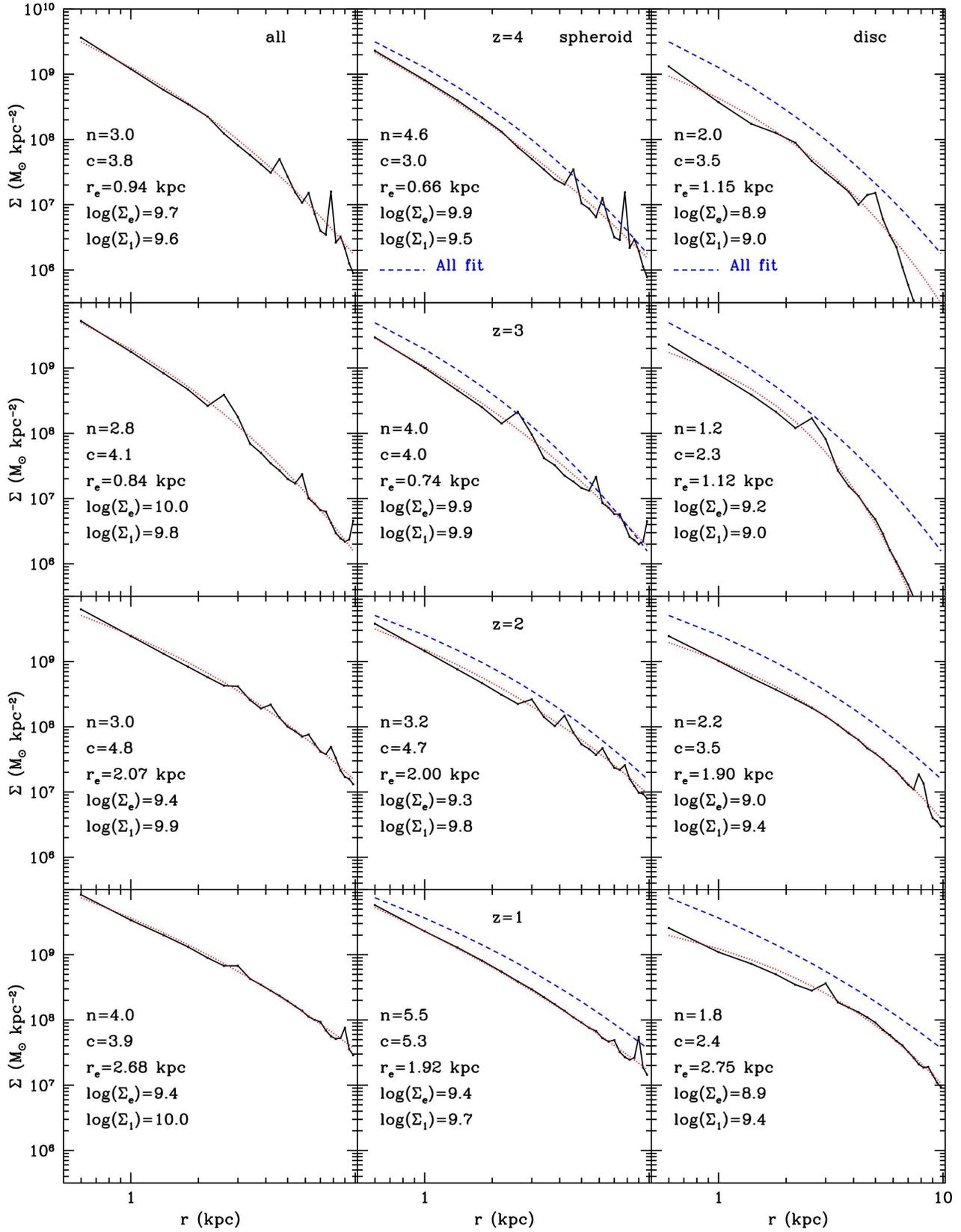}
\caption{Stellar surface density profile of \VLsix \ 
for all (left), spheroid (middle) and disc (right) at redshifts z=4,3,2,1 (from top to bottom). The best-fit Sersic profiles are shown (dotted red curves), together with the sersic index, the concentration parameter, $\Sigma_e$ and $\Sigma_1$. 
 The profile of the spheroid has a high sersic index, n=3-5, at all times, consistent with a classical de-Vaucouleurs-like spheroid.}
\label{fig:VL06BD}
\end{figure*}

\subsection{Stellar surface density profiles and sersic fitting}
\label{sec:profiles}

In this section, we are going to use the stellar surface density profiles from the face-on view of \VLsix \  to study the degree of compactness, as well as the size growth of this galaxy. 
These morphological properties may give us clues about the processes involved in the formation and growth of the galaxy as a whole, as well as the spheroid and the disc separately. 
The relative contribution of these two components determines the final shape, and size of the combined system. 

We describe the profiles of stellar surface density in the face-on view by the Sersic profile \citep{Sersic}:
\begin{equation} 
{\rm ln} (\Sigma) = C + b(n)  [1-(r/r_e)^{1/n}] 
\label{eq:sersic}
\end{equation}
where $C$ is a normalization constant, $\re$ is the effective radius that contains 50\% of the mass, and $n$ is the sersic index that controls the shape of the profile.
Profiles with low values of the sersic index are shallower inside an effective radius than profiles with a high sersic index. 
A pure exponential disc, $n=1$, is an example of a shallow profile. 
A sersic index $n=4$ corresponds to the de-Vaucouleurs profile, typical of spheroids and classical bulges. 
Pseudobulges, on the other hand, have less-concentrated profiles, typically with a low sersic index, $n<2$ \citep{Kormendy04,Laurikainen07,FisherDrory08,Gadotti09}.
The sersic fitting of the profiles is done using a Monte Carlo, least-squared fitting  that samples the parameter space randomly and uniformly, so the probability that the solution converges to wrong secondary minima is low.
Each radial bin has equal weight in the fitting.
In some cases, a neighboring galaxy at large radii produces a feature in the profile that complicates the fitting. In these cases, the profile fitting stops at a radius immediately before that feature. Otherwise, the fitting extends to 10 \kpc \ of projected radius or to a minimum density of log$(\Sigma / \msun \kpc^{-2})=6$, whatever happens first. 

A complementary, non-parametric indicator of the shape of the profile is the concentration. It is usually defined as the ratio of two radii that contains different fractions of the total mass. We define the concentration as:
\begin{equation} 
c = \frac{r_{90}}{r_{50}},
\label{eq:c}
\end{equation}
where $r_{90}$ is the radius that contains 90\% of the mass within a projected radius of 10 kpc and  $r_{50}$ contains 50\%.
Finally, the mean surface density inside a given radius is an indicator of the overall normalization of the profile.
We use $\SigmaOneKpc=M(r<1 \kpc)/\pi$  and $\Sigma_e=\Ms / (2 \pi \re^2)$, as the mean surface density within $1 \kpc$ and $\re$ respectively.


\Fig{VL06BD} shows the face-on profiles of stellar surface density of galaxy \VLsix \ for the disc (right), the spheroid (middle) and the total mass (left) at different redshifts.
At $z=4$, the total profile is steep, with a high sersic index of $n\simeq3$. 
The spheroid component has a much steeper profile, $n\simeq5$ and the disc has a  shallower profile, $n\simeq2$.
Naturally, the composite profile has a sersic index with a value in between the spheroid and disc values.
As the galaxy grows, both spheroid and disc grow in mass and size, although the shape of their profiles remains roughly constant.
The spheroid remains with a typical de-Vaucouleurs profile, $n\simeq4$ and the disc keeps a  shallower profile, $n\le2$.

As the galaxy mass grows by a factor $\sim3.5$ between $z=4$ and $z=1$, the effective radius also grows by a factor $\sim2.8$ during the same period. Therefore, the average surface density inside the effective radius decreases by a factor $\sim2.2$, because the galaxy effective area, $\pi \re^2$, grows faster than the galaxy mass.
The surface density within 1 kpc, $\SigmaOneKpc$, on the other hand, grows monotonically with time, as the mass inside 1 $\kpc$ grows by a factor $\sim2.5$.


To conclude, the profile of the disc component is in general well fitted by a sersic profile with a low sersic index, similar to an exponential profile, $n=1-2$, and low concentration values, $c=2-3$.
The stars with high angular momentum and thick-disc appearance turn out to form a disc with an exponential profile.
In contrast, the spheroid is more concentrated, $c=4-5$ as it has a high sersic index, $n=3-5$, similar to the 
values of local spheroids and classical bulges.
This spheroid formed by VDI or by mergers has a classical, de-Vaucouleurs-like, profile.
This is as opposed to the pseudo-bulges that result from disk instability that is more secular, less violent, occurring at low redshift.
The combined profile has intermediate characteristics, depending on the relative contribution of the disc and spheroid components. The spheroid usually dominates the total mass, so the combined profile, especially toward $z \sim 1$,
resembles the profile of a classical spheroid.

\section{Spheroid growth in a sample of galaxies}
\label{sec:sample}

\begin{figure} 
\includegraphics[width =0.49 \textwidth]{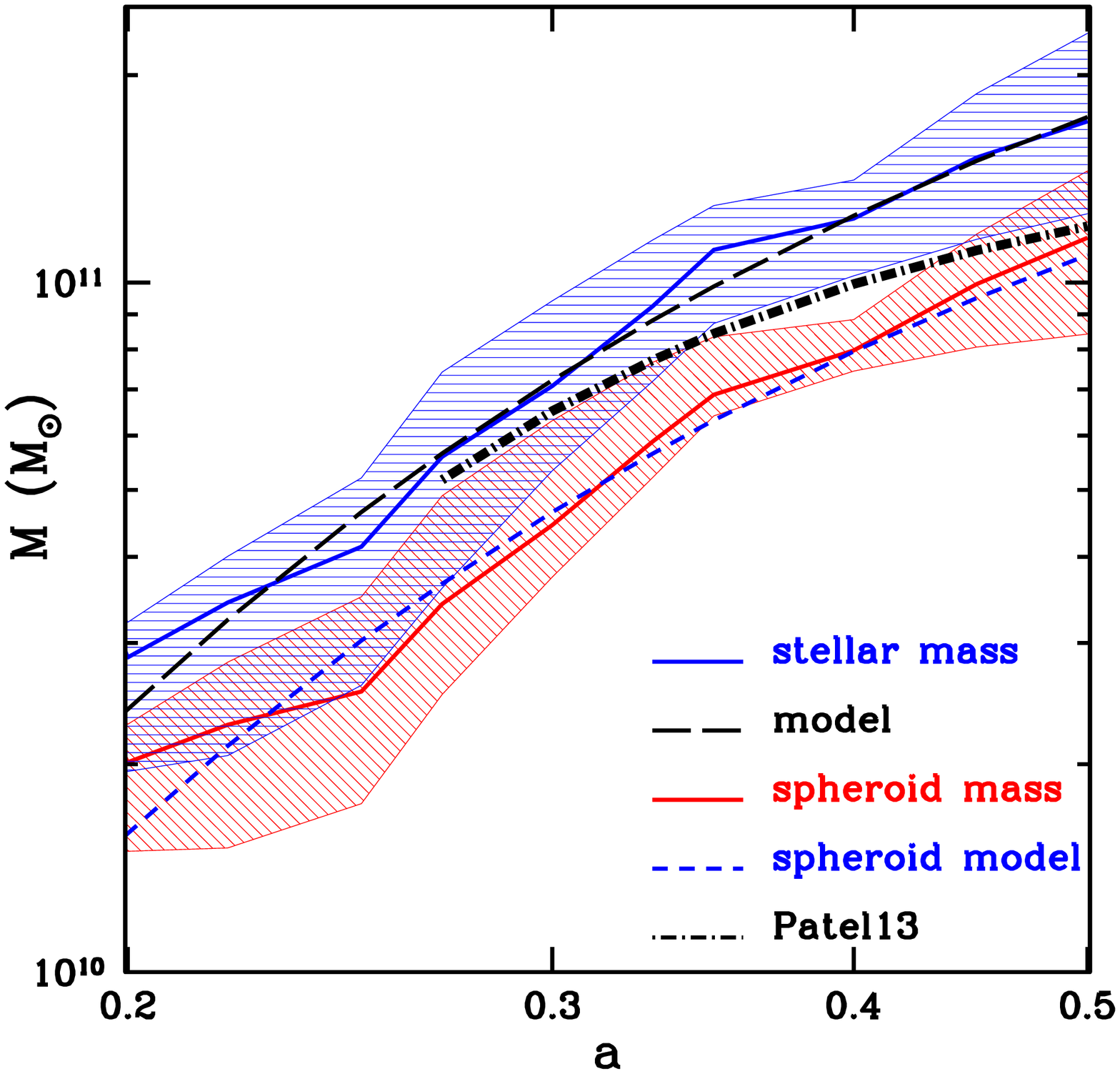}
\includegraphics[width =0.49 \textwidth]{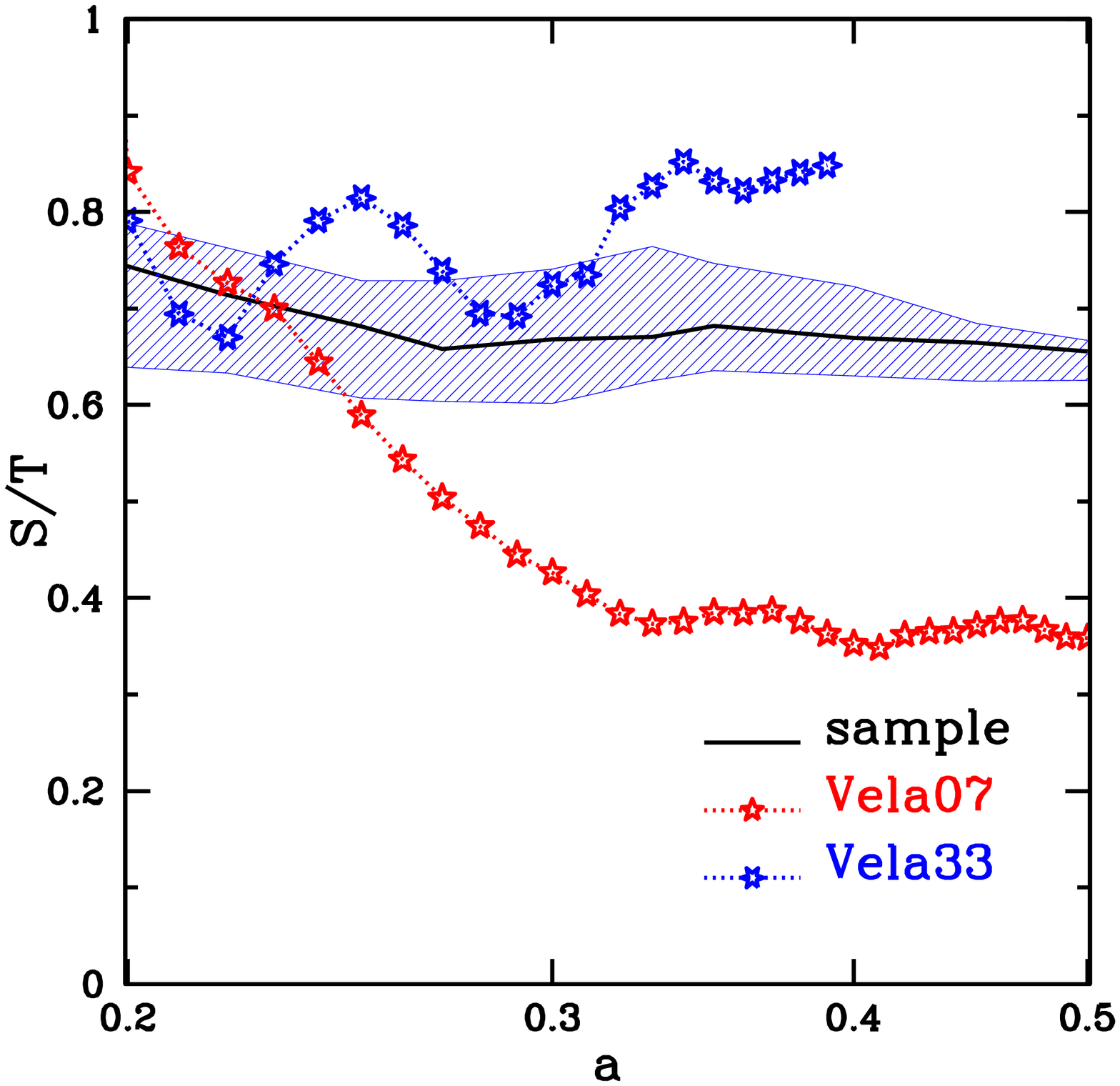}
\caption{Top: Evolution of  total stellar mass (blue) and  spheroid mass (red).
The solid lines correspond to the median of the sample (\VLone \ to \SFGfive) and the shaded regions show the 25\% and 75\% percentiles. 
The dashed lines show the predictions from the toy model, \equ{Mh}
The  dash-dotted line is the observed fit in \citet{Patel13}.
Bottom: Evolution of the spheroid mass fraction. The median is consistent with a time-independent spheroid fraction, as predicted in a VDI cosmological steady state.
High-resolution runs (Vela07 and Vela33) also show periods with a roughly constant spheroid fraction.} 
\label{fig:Mtracks}
\end{figure}

In this section, we generalize some of the results of the previous section, using a sample of zoom-in simulations of halos within a narrow range of virial masses at $z=1$: $\Mv =2-3 \times 10^{12} \ \Msun$ (\VLone \ to \SFGfive), plus a few examples of more massive halos (\SFGtwo, \SFGseven, \SFGeight, \SFGnine).
This allows us to make robust, statistically significant predictions of the characteristic morphological properties of a representative sample of massive galaxies at high redshift, $4\leq z \leq 1$, and their progenitors.

The top panel of \Fig{Mtracks} shows the evolution of the stellar mass and the spheroid mass of all galaxies in the sample from $z=4$ to $z=1$. 
Only halos with a virial mass higher than $10^{11} \Msun$ were considered for analysis. Lower mass halos contain a small number of resolution elements, as well as star particles. Therefore the stellar surface density profiles become too noisy and the morphological properties, such as sersic index and effective radius, get unreliable.
The median of the sample ranges from a stellar mass of $3 \times 10^{10} \Msun$ at $z=4$ to  $\Ms \simeq 2 \times 10^{11} \Msun$ at $z=1$.
Therefore, these galaxies have experienced a dramatic mass build-up by a factor $\sim$5-10 from $z\simeq4$ to $z\simeq1$. 

This mass growth is similar to the observed mass growth in a population of galaxies at different redshifts selected to have a  fixed comoving number density of $n_c=1.4 \times 10^{-4} \Mpc^{-3}$ \citep{Patel13}, 
 a selection that is believed to mimic the evolution of a given sample of galaxies, ignoring major mergers \citep{vDokkum10}.
\citet{Patel13} reported a mass growth by a factor $\sim$2 between $z\simeq3$ and $z\simeq1$.
\citet{vDokkum10} found a similar growth in another sample of slightly more massive, and better resolved, galaxies.
Therefore, observations suggest a significant and continuous mass growth for the high-z progenitors of massive galaxies before $z\simeq1$.

%
%

We can fit the stellar mass growth using the same functional form used for the halo mass growth, \equ{Mh}, and the best fit gives $\alpha=-0.66$ ($s = 0.025  \Gyr^{-1}$) and $M(z=1)=1.7 \times 10^{11} \ \msun$.
This time-scale is very similar to the halo growth, due to the fact that the stellar-to-virial mass ratio is almost time-independent in this mass and redshift range.
This fraction could evolve by a factor of $\sim$2-3 if feedback is stronger in suppressing early star formation in small halos at high redshift, $z=3-4$ \citep{Ceverino13}. 
This deviation is still small compared to the factor $\sim$10 growth in stellar mass between $z=4$ and $z=1$.

%
%

The top panel of \Fig{Mtracks} also shows the growth history of the spheroidal component. It is remarkably similar to the  growth of the total stellar mass.  The median of the sample is continuously growing. This indicates that the major source of spheroid growth is VDI and minor mergers rather than major mergers.
The same functional form, used for the total mass growth, gives
$\alpha=-0.64$ ($s = 0.024  \Gyr^{-1}$) and $M(z=1)=1.1 \times 10^{11} \ \msun$.
This time-scale is very similar to the corresponding time-scale for the total stellar mass. This implies that the spheroid mass is an important component of the total mass. The ratio of the normalization for the spheroid and total mass gives a spheroid mass fraction of $S/T\simeq0.65$. 

The bottom panel of \Fig{Mtracks} shows the evolution of the spheroid-to-total stellar mass fraction of the sample.
The fraction of mass in the spheroid is between 50\% and 90\% of the total mass. Therefore, these massive galaxies are spheroid-dominated. 
The median evolves very slowly from $S/T\simeq0.75$ at $z=4$ to $S/T\simeq0.65$ at $z=1$.
This almost insignificant evolution is consistent with the co-evolution of the disc and the spheroid.
Both disc and spheroid grow in mass at the same rate, as mentioned above. This is predicted in the steady state solution of VDI (DSC), where smooth gas accretion, a marginally unstable disc and the mass migration to the spheroid keep the galaxy in a steady configuration with the disc fraction (disc mass over the total mass inside the disc radius) roughly constant during several Gyrs.

%
%

There are no disc-dominated galaxies ($S/T<0.5$) within our sample of massive galaxies. Such a massive ($\Ms\simeq10^{11}\ \msun$), disc-dominated galaxy is highly unstable and transient, because a massive, gas-rich disc breaks down and  it evacuates material to the center. This material  contributes to the spheroid growth. Therefore, massive, disc-dominated galaxies may correspond to a transient phase in the early evolution of massive spheroids \citep{Genzel08, Forster11a}. 
This first, transient phase is not reproduced in the simulations, probably because the too-weak feedback and too-high star formation efficiency are incapable of preventing the formation of a spheroid at early times.

The suppression of early star formation leads to more disc-dominated systems \citep{Governato10}.
If star formation is inefficient in the small progenitors of these massive galaxies, gas accumulates in a small and stable disc.
Mergers do not contribute much to the spheroid growth because 
 massive outflows prevent large starbursts in the center of small galaxies.
Due to the continuous gas accretion, the disc grows until it is massive enough to become marginally unstable and the spheroid starts to grow and the spheroid fraction increases.
After this transient phase, the disc enters into the steady state solution of VDI, which is the main focus of this paper.
It is possible that these simulations enter the VDI phase too early or too fast, so that the typical spheroid mass fraction is too high.

Runs with higher resolution, stronger feedback and lower star-formation efficiency show two different scenarios. 
Vela33 always has a relatively high spheroid fraction, $S/T=0.7-0.8$, similar to previous simulations.
However, Vela07 grows a disc between $z=4$ and $z=2$, until the spheroid fraction reaches a value of $S/T\simeq0.4$. This value remains constant for the rest of the evolution, as predicted by VDI. 
It seems that  disc-dominated galaxies can grow at high-z due to the inefficiency of early star formation and stronger feedback, as described above. Therefore, the range of $S/T$ values may be higher than in the original sample, but all runs show a period in which the spheroid fraction remains constant, as predicted by VDI.
In addition to stellar feedback, 
other forms of feedback, more effective in regions of high density of baryonic material, such as the center of spheroids, may be needed to efficiently eject star-forming gas and quench star formation in the center, 
in order to suppress the bulge growth. 
AGN feedback may be a good candidate.


\subsection{VDI-driven and merger-driven spheroids}
\label{sec:edgeOn}

\begin{figure} 
\includegraphics[width =0.49 \textwidth]{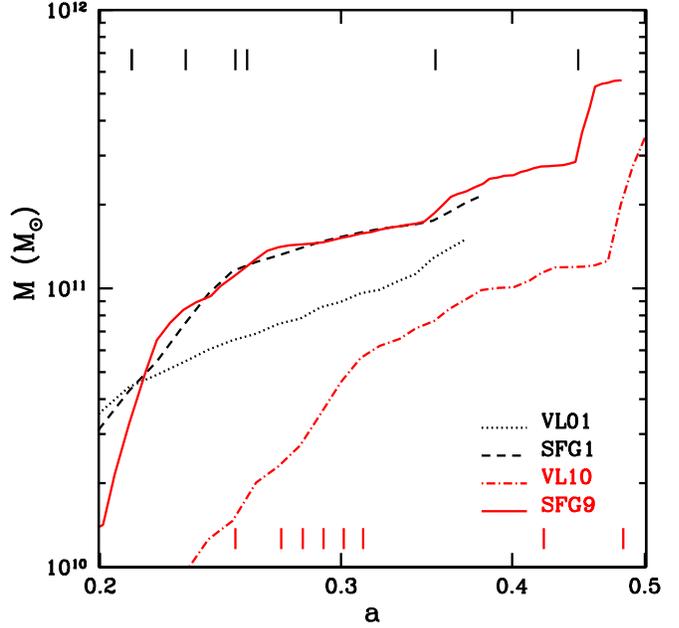}
\caption{ Examples of stellar mass growth dominated by VDI (\VLone \ and \SFGone) or by mergers (\VLten \ and \SFGnine). 
In the first case, less than 10\% of the final stellar mass came in as mergers. In the second case, jumps due to mergers account for 50\%-60\% of the stellar mass growth. 
Black (red) marks in the top (bottom) indicate mergers with a mass ratio higher than 0.1 for \SFGnine \ (\VLten).}
\label{fig:Msamples}
\end{figure}

\begin{figure*} 
\includegraphics[width = \textwidth]{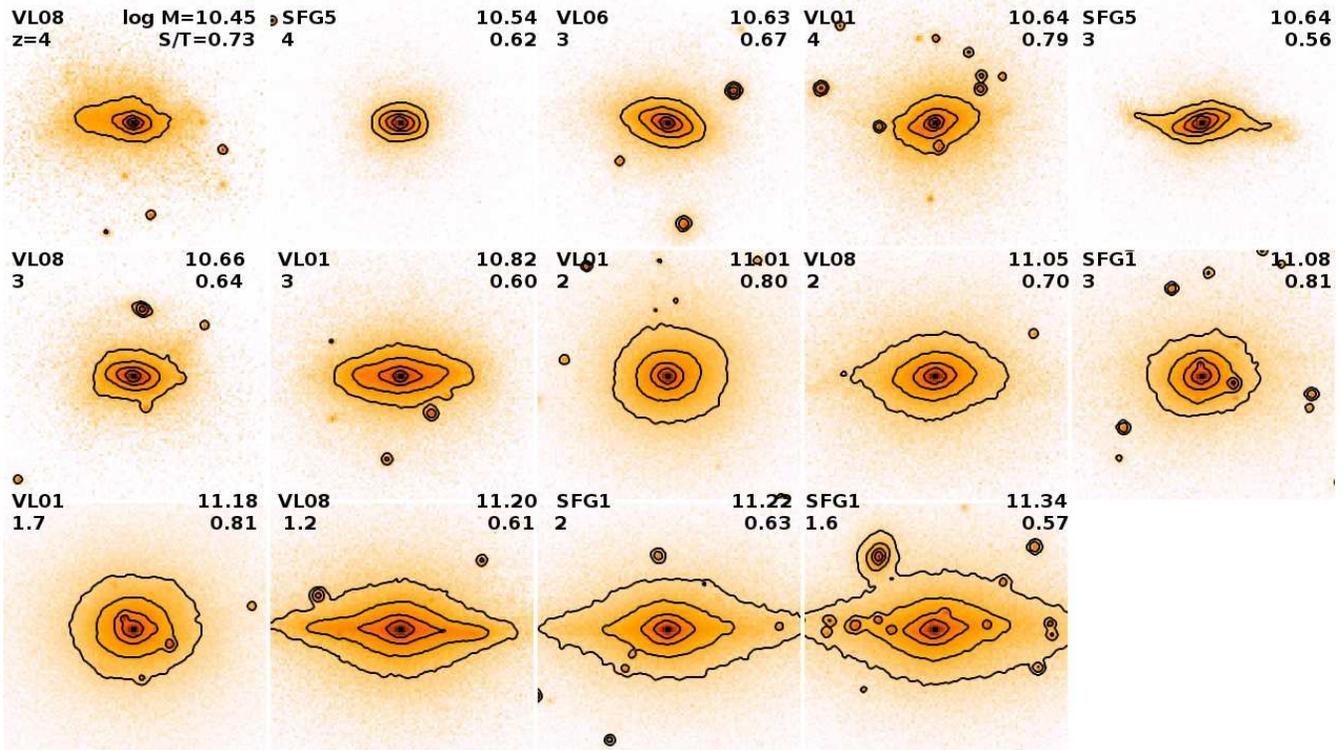}
\caption{Edge-on views of galaxies with a very quiet major and minor merger history, ordered by increasing mass. 
These are galaxies in which the spheroid is mostly growing by VDI (less than 10\% of the galaxy mass came in as mergers of mass ratio higher than 1:10). The spheroids account for $\simeq65\%$ of the galaxy mass and they have classical, de-Vacouleurs profiles, $n=4.7 \pm 1.3$.
The color scale and the contours are as in \Fig{VL06}. In each thumbnail image, the run and the redshift  are indicated on the upper-left corner and the stellar mass and the spheroid fraction are shown in the upper-right corner. 
}
\label{fig:lowmu}
\end{figure*}

\begin{figure*} 
\includegraphics[width = \textwidth]{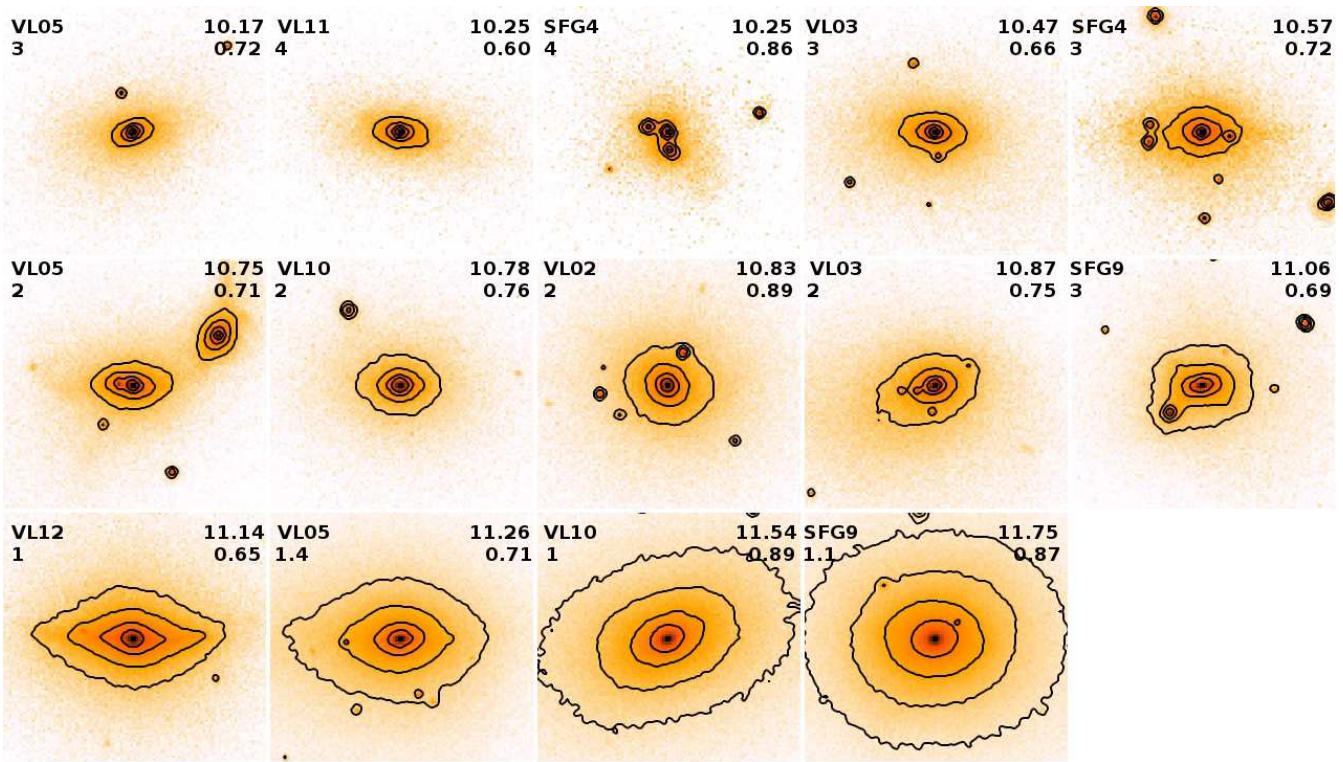}
\caption{Edge-on views of galaxies with a history of minor or major mergers, ordered by increasing mass. 
They are mostly remnants of mergers, which account for more than 40\% of the galaxy mass.
The spheroid mass fraction and the profiles, are all similar to the galaxies with a quiet merger history.}
\label{fig:highmu}
\end{figure*}

\begin{figure*} 
\includegraphics[width = \textwidth]{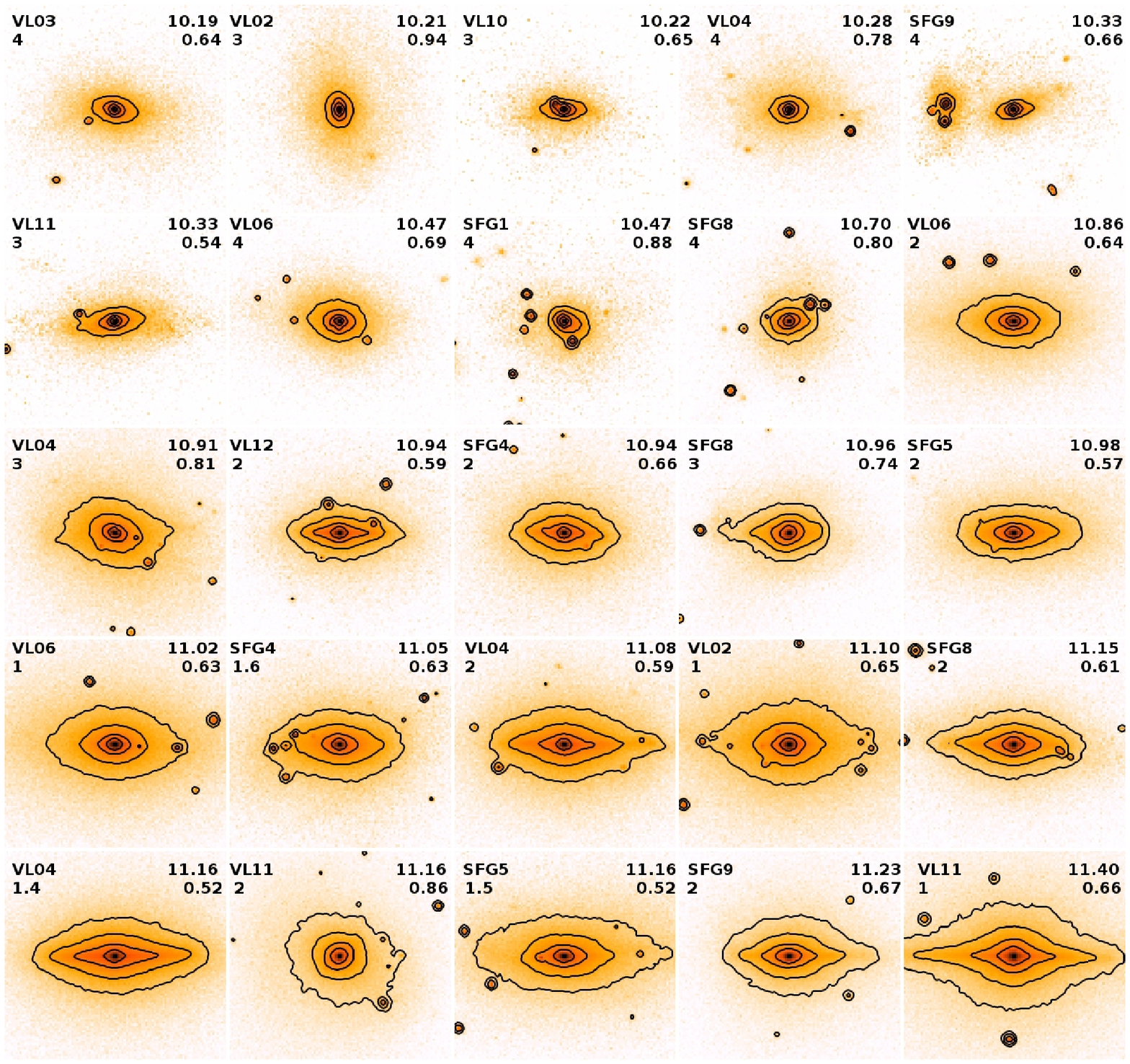}
\caption{Edge-on views of galaxies with intermediate values of $\mu_{10}$, ordered by increasing mass. 
 These are galaxies with a mixed morphology and their spheroids may be formed by both VDI and mergers.
 They have similar values for spheroid fraction and sersic index than the VDI-driven sample.
  The color scale, the contours and the values are as in \Fig{lowmu}.}
\label{fig:middlemu}
\end{figure*}

One way to distinguish between a spheroid growth due to VDI or mergers is by utilizing merger histories.
 The mass fraction that came in through mergers with a mass ratio above a given value (1:10 for example) serves as a measure of the importance of
mergers in the galaxy (and spheroid) growth. Galaxy merger trees were constructed using a modified version of the structure
finder \textsc{AdaptaHOP} \citep{Tweed09}. 

The merger tree is constructed from the star particles only.
First, a galaxy catalog is constructed at each snapshot.
Then, galaxies from two consecutive snapshots are associated with each other based on the total mass of particles that they share. For each galaxy at a given snapshot, one descendant galaxy is assigned in the subsequent snapshot, and one or more progenitors are defined in the preceding snapshot. For each merger, corresponding to a case of multiple progenitors to a given galaxy, the merger mass ratio is defined by the stellar masses of the
main (most massive) progenitor and the most massive among the secondary progenitors. The parameter $\mu_{10}$ is defined for each galaxy at each snapshot as the fraction of its mass that has been added to the galaxy during the history of its main progenitor  through mergers of  mass ratio larger than 1:10, namely as minor or major mergers. More details can be found in Tweed et al. (in preparation).

A low value of $\mu_{10}$ indicates a galaxy growth dominated by in-situ star formation. For example, a galaxy with $\mu_{10}=0.1$ means that only 10\% of its current stellar mass has been formed in other distinct galaxies and came in
through mergers with a mass ratio higher than 0.1.
%
\Fig{Msamples} shows several examples of quasi-continuous galaxy growth (\VLone \ and \SFGone), as well as merger-dominated growth (\VLten \ and \SFGnine).
\SFGone \ ($\mu_{10}=0.06$ at $z$=1.6) and \VLone \ ($\mu_{10}=0.08$ at $z$=1.7) are examples in which VDI is the main process of spheroid growth.
For the merger-dominated histories, the $\mu_{10}$ values are much higher.
In the case of \SFGnine, half of the final stellar mass at $z$=1.1 came by mergers ($\mu_{10}=0.50$). 
For \VLten, 64\% of the galaxy growth at $z$=1 was dominated by jumps due to mergers.

%
We select all galaxies at 4 different redshifts, $z=$4,3,2 and the last available snapshot, $z\simeq 1$, with the exception of the most massive runs (\SFGtwo \ and \SFGseven) that finished at high-z.
We divide the snapshots into three subsamples according to their $\mu_{10}$ values.
The first subsample includes all cases within the first quartile of the $\mu_{10}$ distribution:
The 25\% of all snapshots with the lowest values of $\mu_{10}\leq0.1$.
The second subsample includes the intermediate values of $\mu_{10}$, around the median, $\bar{\mu}_{10}=0.26$ (second and third quartiles).
Finally, the third subsample includes the forth quartile, with the highest values of $\mu_{10}>0.4$.

%
The edge-on views of the stellar mass surface density of the galaxies in the first quartile of $\mu_{10}$ are shown in  \Fig{lowmu}. 
They represent the cases in which VDI is the dominant mechanism of spheroid growth.
The spheroid mass fraction is high, $S/T=0.65 \pm 0.09$, similar to the median value of the whole sample.
These VDI-driven spheroids also have a typical de-Vacouleurs profile with a high sersic index, $n=4.7 \pm 1.3$.
Therefore we conclude that VDI produces spheroids with classical profiles.
The edge-on views also show some cases with extended discs, especially for relatively low spheroid fractions, $S/T\simeq0.6$, and low redshifts, $z<2$, as well as mostly round objects, more common at $z\geq2$.
These round galaxies tend to be small, with an effective radius of $\re=1.0 \pm 0.5$ kpc.
The more extended galaxies at $z\simeq 1-2$ are twice larger,  with an effective radius of $\re=2.1 \pm 0.7$ kpc.

%
The galaxies within the forth quartile of $\mu_{10}$ have their mass accretion history dominated by mergers.
These merger remnants tend to be spherical (\Fig{highmu}) and they have high spheroid fractions, $S/T=0.72 \pm 0.09$, slightly higher than the VDI-driven examples.
Their spheroids have profiles with a sersic index of $n=4.3 \pm 1.6$, consistent with the classical de-Vacouleurs profile. 
 The merger remnants at $z>2$ tend to be small and compact, with an effective radius of $\re=1.7 \pm 0.8$ kpc. 
 This compactness is a direct consequence of the dissipative nature of the mostly wet mergers at high-z.
 The remnants at lower redshifts,  $z\simeq 1-2$, are larger ($\re=2.2 \pm 0.6 \kpc$).
 This is probably due to the lower gas fractions of the progenitors, which makes the merger less dissipative than their high-z counterparts.
 
%
The galaxies with intermediate values of $\mu_{10}$ (second and third quartiles) are shown in \Fig{middlemu}. 
This population is a mixture of round, compact spheroids and more flattened galaxies.
A disc component is visible in some cases.
The spheroid fraction, $S/T=0.65 \pm 0.11$, is similar to the VDI-driven examples. 
The same is true for the sersic index of the spheroid, $n=4.7 \pm 1.3$. 
It seems that the galaxies with intermediate values of $\mu_{10}$ have properties similar to the VDI-driven galaxies with low  $\mu_{10}$. 
However, due to their relatively high values of $\mu_{10}$, 
the spheroid growth may be dominated by both mergers and VDI. 
A full dissection of the spheroid growth in these mixed cases turns out to be a rather complex task.

%
We conclude that the pure VDI-driven spheroids are very similar to the merger remnants in terms of overall morphology.
They both have a high sersic index, consistent with a classical de-Vacouleurs profile. 
The mean surface density inside the half-mass radius have the same median in both samples, $\Sigma_e=4 \pm 3  \times 10^9  \ \msun \kpc^{-2}$. This is because the two processes are highly dissipative and they both produce compact spheroids, 
with the violent merger history of the remnants leading to a slightly higher effective radius and spheroid mass fraction.

\section{Classical spheroids and exponential discs}
\label{sec:morpho}

\begin{figure}
\includegraphics[width =0.49  \textwidth]{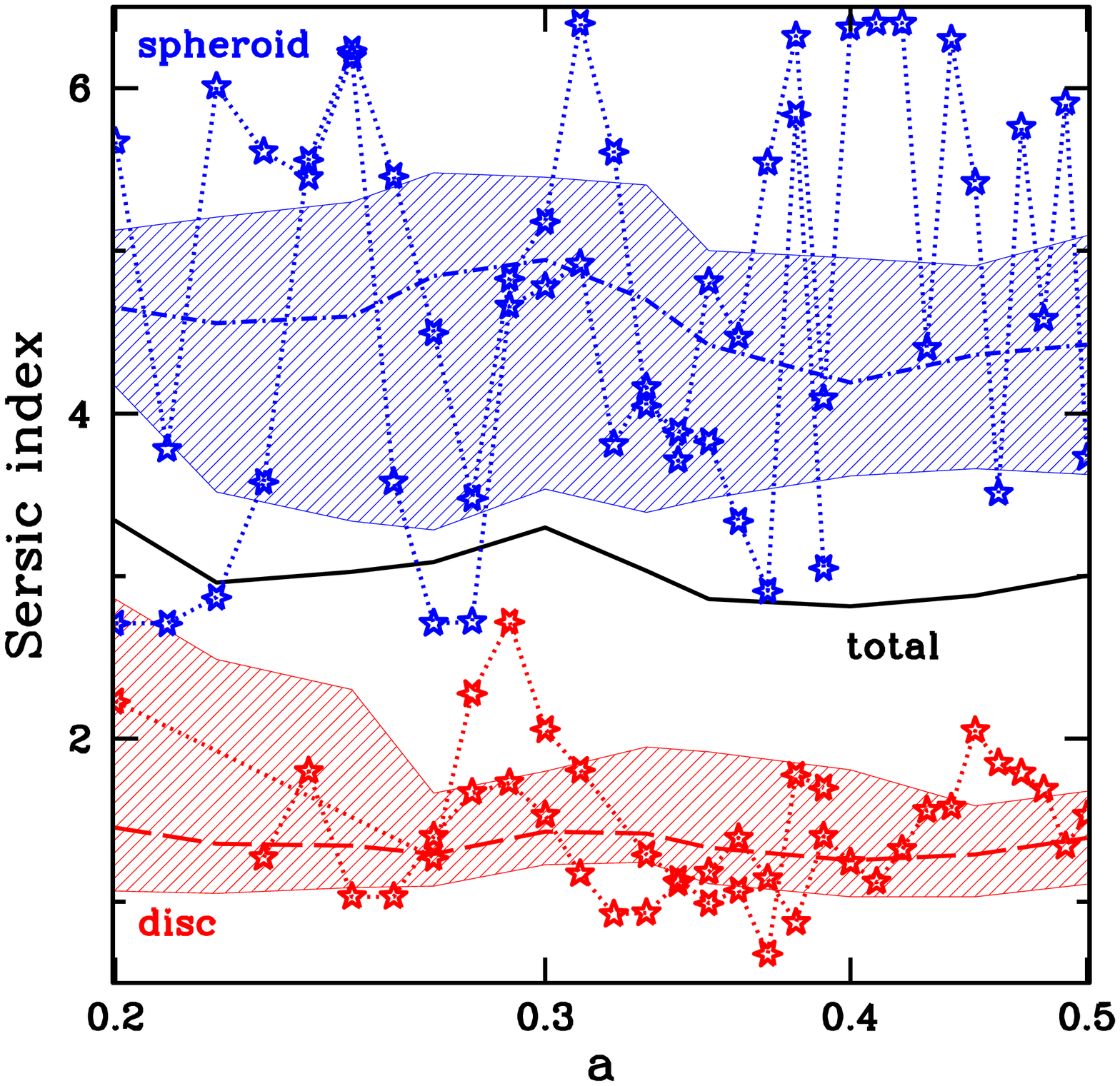}
\includegraphics[width =0.49 \textwidth]{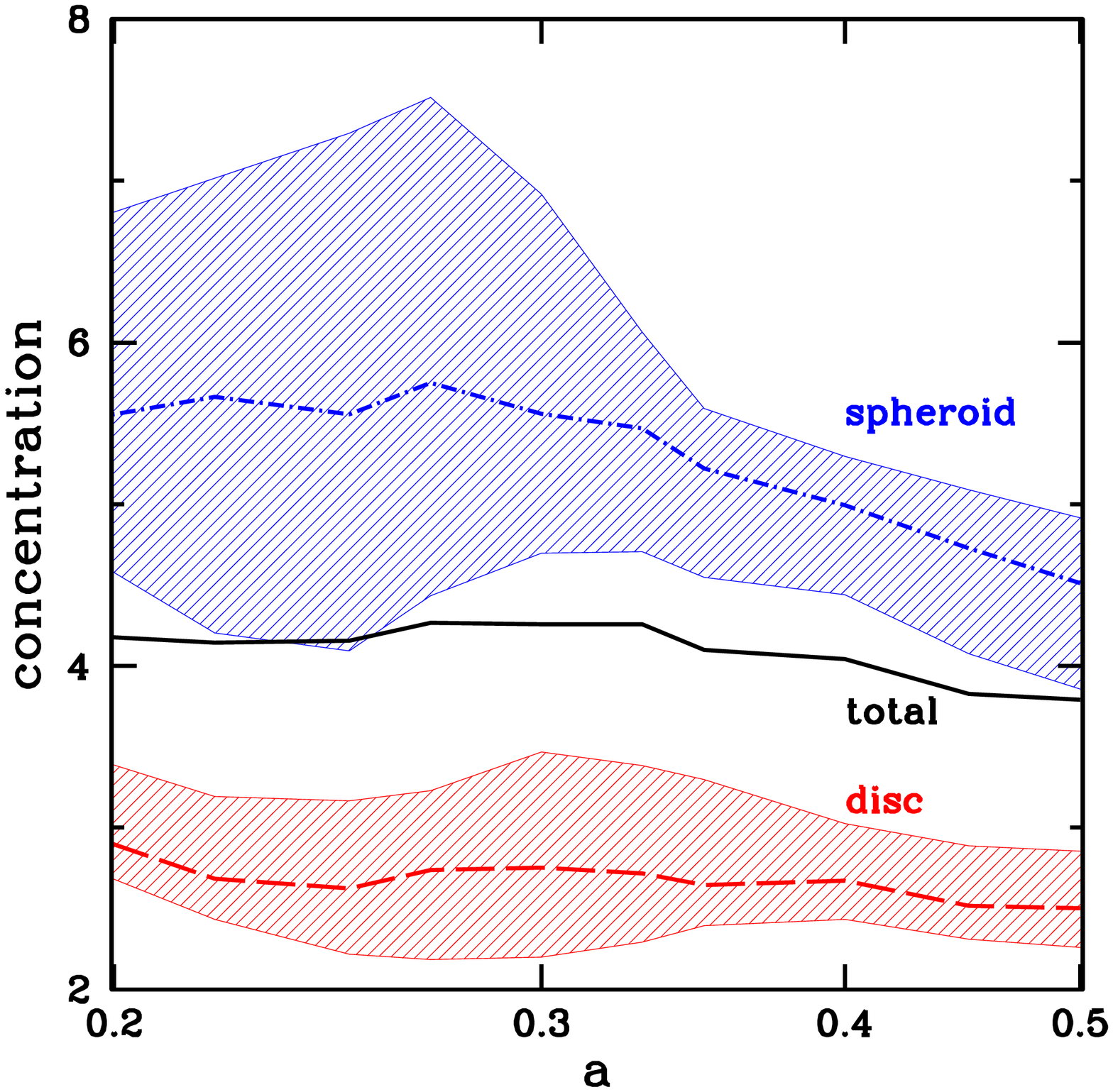}
\caption{Evolution of the sersic index, $n$ (top), and concentration, $c$, (bottom) for the different components.
The sersic index of the  combined profile remains constant around $n\simeq3$.
The spheroid component has a profile close to the  de-Vaucouleurs profile, $n=4$ and the disc is close to exponential, $n=1$.
This result is independent of feedback, resolution or star-formation efficiency, as shown in Vela07 and Vela33, represented as dotted lines, as in \Fig{Mtracks}.
The concentration of the total profile remains constant around $c\simeq4$.
The concentration of the spheroid is high, close to the concentration of the de-Vaucouleurs profile, $c=5.5$.
The concentration of the disc is much lower, similar to the concentration of the exponential profile, $c=2.3$.}
\label{fig:nall}
\end{figure}

At all redshifts, the median sersic index of the spheroid profile is always high, close to $n\simeq4-5$ with a scatter of about unity (\Fig{nall}), namely, all the galaxies in our sample have classical spheroids.
At the same time, the component with high angular-momentum, defined as the disc, has a low sersic index and a close-to exponential profile.
Its mean sersic index is always $n\simeq1-2$, independent of time, with a scatter of $\pm 0.5$,
excluding a few cases of major merger remnants that have very little material with high $\fj$ values, especially at high-z.

At z=1, there are examples of very compact and steep-profile spheroids, like \VLsix, with $n=5.5$, shown in \Fig{VL06BD}, as well as examples of relatively shallow spheroidal profiles, like \VLfive, with $n=2.9$.
However, there is no one single case of low sersic index, $n\le2$, more typical of pseudobulges,
as being produced by more secular processes \citep{Kormendy04}.

The median of the sersic index of the combined profile is about $n\simeq3$ at all times, indicating an overall spheroid-like profile.
The $n<2$ profiles found in some galaxies, such as \VLeight \ at $z=1$, are actually due to the combination of a classical spheroid ($n>3$), with a relatively massive exponential disc ($n\simeq1$), which yields a relatively low spheroid fraction, $S/T=0.6$.


The concentration (\Fig{nall}) has a similar behavior to the sersic index.
The concentration of the spheroidal component is always high, $c=5 \pm 2$, and the median of the sample coincides with the concentration of the  de-Vaucouleurs profile ($c=5.5$).
On the other hand, the disc component is less concentrated, $c=2.7 \pm 0.7$, similar to the concentration of the exponential profile  ($c=2.3$).
The total profile have always a high concentration,  $c=4.1 \pm 0.8$, closer to the concentration of the spheroid. This is due to the relatively high spheroid mass fractions within the sample. 


There is no clear systematic evolution of the sersic index and concentration with time, although the galaxies grow by a factor of $\sim10$ between $z=4$ and $z=1$. This generalizes the result from the example discussed in section  \se{BD}, where $n=3-4$ at all times.
The shape of the combined profile does not evolve because it depends 
on the intrinsic shapes of its components (spheroid and disc), as well as their relative masses and sizes
and these properties are largely independent of time in the redshift range explored here.


The scatter in the sersic index of the total profile is large, especially at high-z. For example \VLone \ has a profile consistent with an exponential disc, $n=1.0$ at $z=3$. At the same redshift, another galaxy of similar mass, \VLfour, has a much steeper profile, $n=4.7$. This diversity is due to the non-negligible range of spheroid-fractions: \VLone \ has $S/T=0.6$ and \VLfour \ has $S/T=0.8$.
A large scatter is also present at $z=1$, where the range of sersic index extends from $n=1.9$ ($S/T=0.5$) for \VLfour \ to $n=5$ ($S/T=0.9$) for \VLten. The edge-on maps shown in \Fig{lowmu}-\ref{fig:middlemu} already showed this large variety of morphologies among massive galaxies at $z=1$.

This large scatter is consistent with the large range of sersic index observed by \citet{vDokkum11} in a sample of massive galaxies at $z=1-1.5$. 
Massive and compact spheroids tend to have a high sersic index, $n\simeq3$ at $z\simeq2$ \citep{vDokkum08, vDokkum10, Cassata10, Szomoru12}.
In the $z\simeq2$ sample of \citet{Patel13}, quiescent galaxies have a high sersic index, $n=3-4$, while star-forming galaxies of similar mass, $\Ms\simeq7 \times 10^{10} \ \msun$, tend to have exponential profiles, $n\simeq1$.
The simulations reported in this paper reproduce this diversity of morphologies, although they miss a significant sample of pure exponential discs that account for $\sim20\%$ of the sample in \citet{vDokkum11} at $z\simeq1$ and about two thirds of the sample of smaller galaxies in \citet{Patel13} and in \citet{Weinzirl11}.


As mentioned above, this lack of disc-dominated galaxies may be due to an overestimate of star-formation efficiency and underestimate of feedback strength in simulations. This causes the formation of an early massive bulge in all cases.
Vela07 is one case of a disc-dominated galaxy, but even in this case, the spheroidal component has a classical, de-Vaucouleurs-like profile, $n=4.8 \pm 1.1$. Therefore, simulations with strong feedback have smaller spheroids 
\citep{Guedes11}, but their structural properties at high-z are similar to the weak-feedback runs.

\section{Mass and Size Growth}
\label{sec:MassSize}

In the previous sections, we saw that the progenitors of massive galaxies grow in mass and size, from small proto-galaxies with stellar masses of a few times $10^{10} \ \msun$ at $z \simeq4$ to more mature galaxies of $\Ms \sim 10^{11} \ \msun$ at $z=1-1.5$.
The mass and size growth described here occurs after the first growth of the spheroidal component, when the spheroid surface density increases due to the 
dissipative shrinkage of gaseous discs into compact star-forming systems -- ``blue nuggets" \citep{DekelBurkert}.
Due to the too-high star-formation efficiency mentioned above, this process happens too early in the simulations 
used here, typically at $z>4$.
In simulations with stronger feedback \citep{Ceverino13}, this  
compaction tends to occur at somewhat later redshifts. We address the compaction process in detail in Zolotov et al. (in prep.).
Here, we focus on the mass-size evolution after this first compaction.

\subsection{Mass-size relation at $z\sim2$ compared to observations}
\label{sec:MassSize2}

\begin{figure} 
\includegraphics[width =0.49 \textwidth]{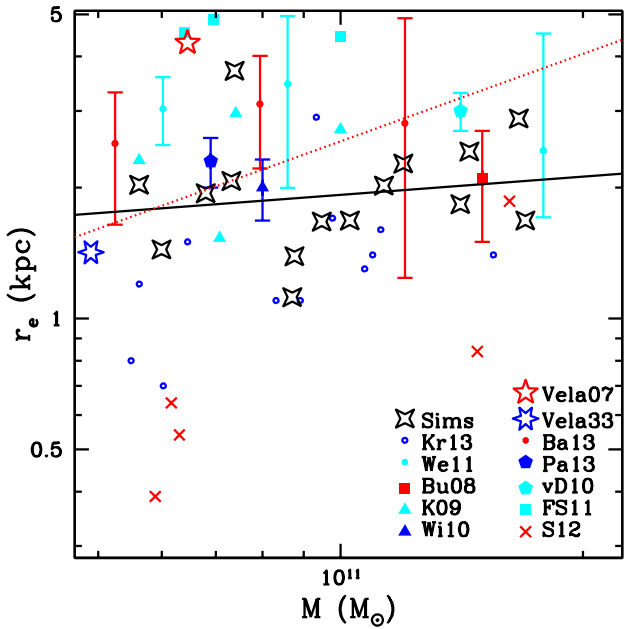}
\caption{Mass-size relation at $z\sim2$ from simulations and observations. 
The black line is the best linear fit to the simulations.
The dotted red line is the threshold used in \citet{Barro13} to separate compact from extended galaxies
in their observed sample at $z$=1-3. Based on this threshold, most of the simulated galaxies are compact, with 3 marginal cases and one extended system.}
\label{fig:ReMs_comp}
\end{figure}

\Fig{ReMs_comp} compares the mass-size relation predicted from the simulations at $z\simeq2$ with a set of $z\simeq2$ observations compiled from the literature. 
In this way we can check the overall agreement with observations and how robust the predictions from these simulations are in the mass-size plane. 
However some caution is needed in this comparison because many of these observed samples are not complete or homogeneous.
As it can be seen from the figure, these simulations roughly agree with several observations of massive galaxies at $z\simeq2$ \citep{Buitrago08, Kriek09a, Williams10, vDokkum10, Weinzirl11, Law12, Barro13,Patel13, Krogager13}.
These simulated galaxies at $z\simeq2$ are already massive ($\Ms>5 \times 10^{10} \ \msun$) and they have an effective radii of $\re= 2 \pm 1 \kpc$ with little mass dependence, consistent with the results in \citet{Law12} and \citet{Morishita14} for a similar mass and redshift range. In addition, most of them are compact,  according to the definition adopted by \cite{Barro13} based in their position in the $\Ms-\re$ diagram. There are 3 marginal cases and one extended system.
 
However, some cases of $z\simeq2$ galaxies are missing from this sample.
For example, none of the simulations show large, disc-dominated galaxies with $\re> 4 \kpc$, as in the sample of \citet{Forster11a}. As mentioned above, stronger feedback may be needed in order to remove gas from the center before it turns into stars and forms a compact and dense bulge \citep{Governato10}. 
For example, the run with stronger radiative feedback, Vela07, shows a disc-dominated galaxy ($n=1.5$) with a mass of $\Ms=6 \times 10^{10} \ \msun$ at $z=2$ and an effective radius of 4 kpc, similar to the observed extended galaxies.
However, the other simulation with strong feedback, Vela33, shows a much smaller effective radius, consistent with the rest of runs. It seems that the addition of radiative feedback increases the range of the sample towards more extended systems but the typical radii are still similar than in the runs with weak feedback and higher star-formation efficiency.

 \begin{figure*} 
\includegraphics[width = \textwidth]{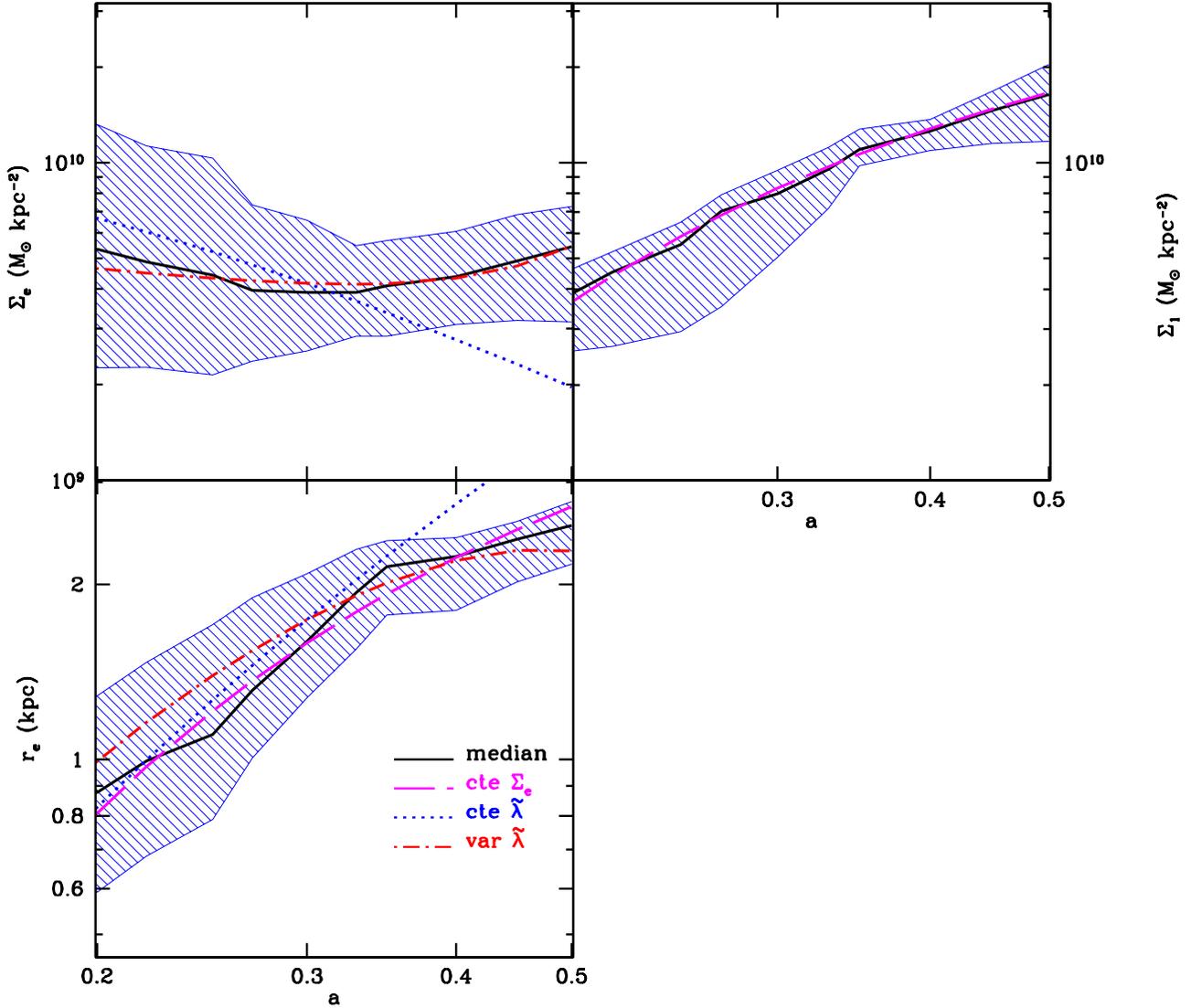}
\caption{Evolution of the mean surface density inside $\re$ (top-left), mean surface density inside 1 kpc (top-right), and the effective radius (bottom-left). 
Galaxies grow in mass and size such that $\Sigma_e \sim$ constant 
and $\SigmaOneKpc$ grows at the same rate as the total mass.}
\label{fig:Sigmae1Re_Evo}
\end{figure*}

In the other extreme of the size range, we do not find any case of ultra-compact ($\re<1 \kpc$) galaxies among our $z=2$ sample. These galaxies are described by \citet{Szomoru12} and they tend to be quiescence and not very massive ($\Ms<10^{11} \ \msun$). It is possible that they stopped forming stars at much earlier redshifts ($z\ge3$), when the typical radius of the galaxies with this mass was much smaller than at $z=2$. As shown next, the simulated galaxies at $z\ge3$ have indeed several examples of these ultra-compact galaxies with $\re<1 \kpc$.
However, these simulations do not have an efficient quenching mechanism, such as radiative or AGN feedback, so the progenitors of these galaxies in the simulations do not quench and continue growing in mass and size. The typical values of the specific star formation rates of these massive galaxies ($\Ms\simeq10^{11} \ \msun$) at $z\simeq1$ are sSFR=$0.16 \pm 0.09 \ {\rm Gyr}^{-1}$. We defer further details about quenching to future work (Zolotov et al., in prep.).


%
\subsection{Evolution of the surface density within $\re$}
\label{sec:SigmasEvo}

The  surface density within the effective radius, $\Sigma_e$, indicates how dense these galaxies are and gives some clues about the dissipative processes that make compact spheroids.
First, we are going to analytically estimate  the surface density within the effective radius of a disc formed by dissipative gas contraction within a dark matter halo \citep{FallEfstathiou80, MoMaoWhite98}. 
We assume that a characteristic disc radius scales with the halo virial radius via a contraction factor, $\tilde{\lambda}$:
\be
r_{\rm d} = \tilde{\lambda} \ r_{\rm v}
\label{eq:rd}
\ee
If there is no significant loss of angular momentum, this contraction factor is related to the halo spin, $\tilde{\lambda}\simeq0.05$ \citep{Bullock01,Dekel13}.
Assuming an exponential disc, the effective radius is related to the disc radius by the concentration, 
$\re= r_{\rm d} / c$, where $c=2.4$.
The spherical collapse model and the virial theorem lead to the following virial relation, approximated in the Einstein-deSitter regime \citep{Dekel13} by
\be
r_{\rm v} = 100 \ {\rm kpc} \ M_{12}^{1/3} (1+z)_3^{-1}
\label{eq:rv}
\ee  
where $M_{12}=\Mv / 10^{12} \ \msun$ and $(1+z)_3=(1+z)/3$.
Using the above equations, the surface density within $\re$ is
\be
\Sigma_e= 3.6 \times 10^9 \ \msun \ {\rm kpc}^{-2} m_{\rm d,0.1} \ \tilde{\lambda}_{0.05}^{-2} \ M_{12}^{1/3} \ (1+z)_3^2 
\label{eq:Sigmae}
\ee
where $\tilde{\lambda}_{0.05}=\tilde{\lambda}/0.05$, and $m_{\rm d,0.1}=m_{\rm d}/0.1$ refers to the baryonic mass relative to $\Mv$ \citep{Dekel13}.

\Fig{Sigmae1Re_Evo} shows $\Sigma_e$, measured from the simulations, as a function of the Universe expansion factor. 
There is little evolution of $\Sigma_e$ for most of the sample.
Between $z=4$ and $z=1$, it is always around a constant value of about, $\Sigma_e\simeq (5 \pm 2) \times 10^9 \ \msun \kpc^{-2}$.
Using \equ{Mh} for the halo mass growth and assuming the fiducial values in \citet{Dekel13}, $\tilde{\lambda}_{0.05}=1$ and $m_{\rm d,0.1}=1$, the toy model predicts a decline in $\Sigma_e$, which seems inconsistent with the simulations. In particular, the simulations are denser by a factor $\sim$2 at $z<2$. 
One possible explanation is that the baryons in the simulations loses angular momentum during its collapse, as described in \citet{Danovich14}.
In fact, \citet{Dekel13} found in the same simulations used here a systematic trend of $\tilde{\lambda}$ with redshift (see their Figure 12). Their measure of $\tilde{\lambda}$ is somewhat higher than the fiducial value, $\tilde{\lambda}=0.06$ at $z$=3-4 and lower, $\tilde{\lambda}=0.03$, at $z\simeq1$. If we fit this trend by
\be
\tilde{\lambda}_{0.05}=-0.66 \ (1+z)_3^{-1} + 1.6
\label{eq:lambda}
\ee
and we allow $\tilde{\lambda}_{0.05}$ to vary accordingly in \equ{Sigmae}, we obtain no evolution of $\Sigma_e$ with time, in  agreement with simulations. 

These results suggest that the progenitors of massive galaxies grow in mass and size between $z=4$ and $z=1$ in such a way that the mean surface density inside the effective radius evolves very slowly with time.
This could be due to a combination of the Universe expansion, which tends to decrease $\Sigma_e$ and dissipative processes,
as well as angular momentum losses,
 which tend to decrease the galaxy-to-halo scale ratio, increasing $\Sigma_e$.

The most massive simulations, such as \SFGnine, also show little evolution of $\Sigma_e$, although their values are significantly higher, $\Sigma_e > 10^{10}  \ \msun \kpc^{-2}$ 
after its first growth at $z>4$.
During this early stage, the surface density increases to this high value due to the dissipative shrinkage of a gaseous disc into a compact spheroid \citep[][Zolotov et al.]{DekelBurkert}. 
After this dissipative disc contraction, the surface density remains high during the later growth.

\subsection{Evolution of the effective radius}
\label{se:Retracks}

\Fig{Sigmae1Re_Evo} also shows the evolution of the effective radius with time.
At $z=4$, the progenitors of massive galaxies have a large diversity of sizes, although they are typically small, with sub-kpc sizes. The median of the effective radius is $\re\simeq0.9 \kpc$.
Two examples with similar stellar mass illustrates this variety: the very extended, $\re\sim2 \kpc$, \VLone \ run and the very compact, $\re \sim 0.8 \kpc$, \SFGeight \ run.
Between $z\simeq3$ and $z\simeq2$, there is a strong size evolution. The effective radius of the whole sample grows from $\re \simeq 1 \kpc$ at $z=3$ to $\re \simeq 2 \kpc$ at $z=2$.
However, this rate decreases at lower redshifts. By $z\simeq1$ the mean effective radius is $\re \simeq 2.5 \kpc$, only 50\% higher than $z=2$. 

We can understand this evolution by using the simple toy model described above.
Using \equ{rd}, \equ{rv} and the definition of concentration, 
\be
\re=2 \ {\rm kpc} \   \tilde{\lambda}_{0.05} \ M_{12}^{1/3}  \ (1+z)_3^{-1}
\ee
Assuming a constant value of $\tilde{\lambda}_{0.05}=1$, the model overestimates the size evolution, when compared to simulations. 
Using the evolution of  $\tilde{\lambda}_{0.05}$ of \equ{lambda} gives a better agreement.

Alternatively, we can describe the size growth by using the fact that galaxies grow in mass and  size such that the stellar surface density inside the effective radius evolves very slowly, $\Sigma_e\sim $ constant. Therefore, $\re \propto \Ms^{1/2}$ and we can use the growth of the stellar mass described in \se{sample},
where $\alpha=-0.4$ and $\re(z=1)= 2.7 \kpc$ give the best fit.
This size growth can also be described by $\re \propto (1+z)^{-1.31}$, which is remarkably similar (1\% difference) to the observed fit, discussed in \citet{vDokkum10}, although their sample have more massive and larger galaxies.

\subsection{Evolution of the surface density within 1 kpc}
\label{sec:Sigma1Evo}

\begin{figure}
\includegraphics[width = 0.5 \textwidth]{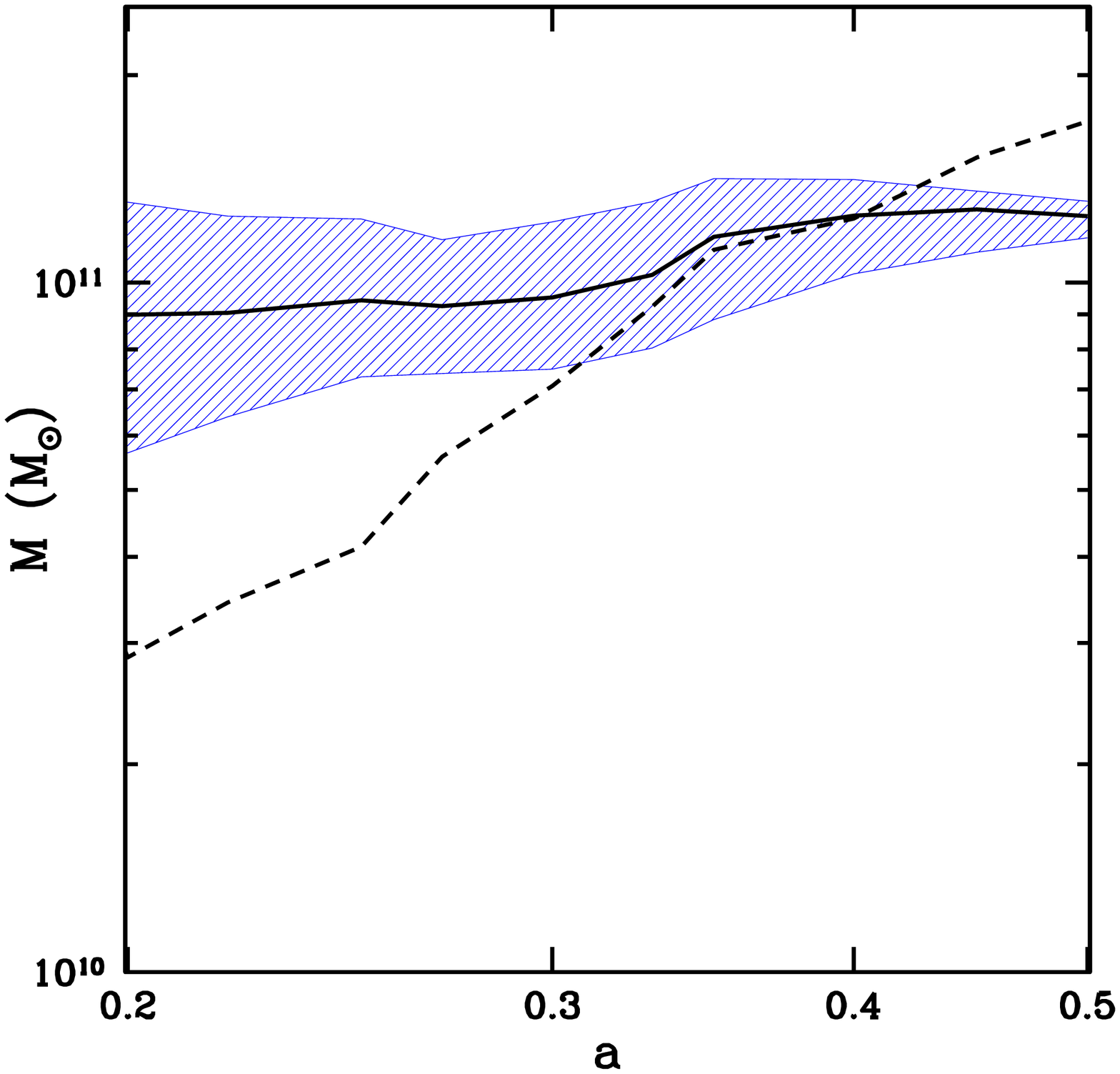}
\includegraphics[width = 0.5 \textwidth]{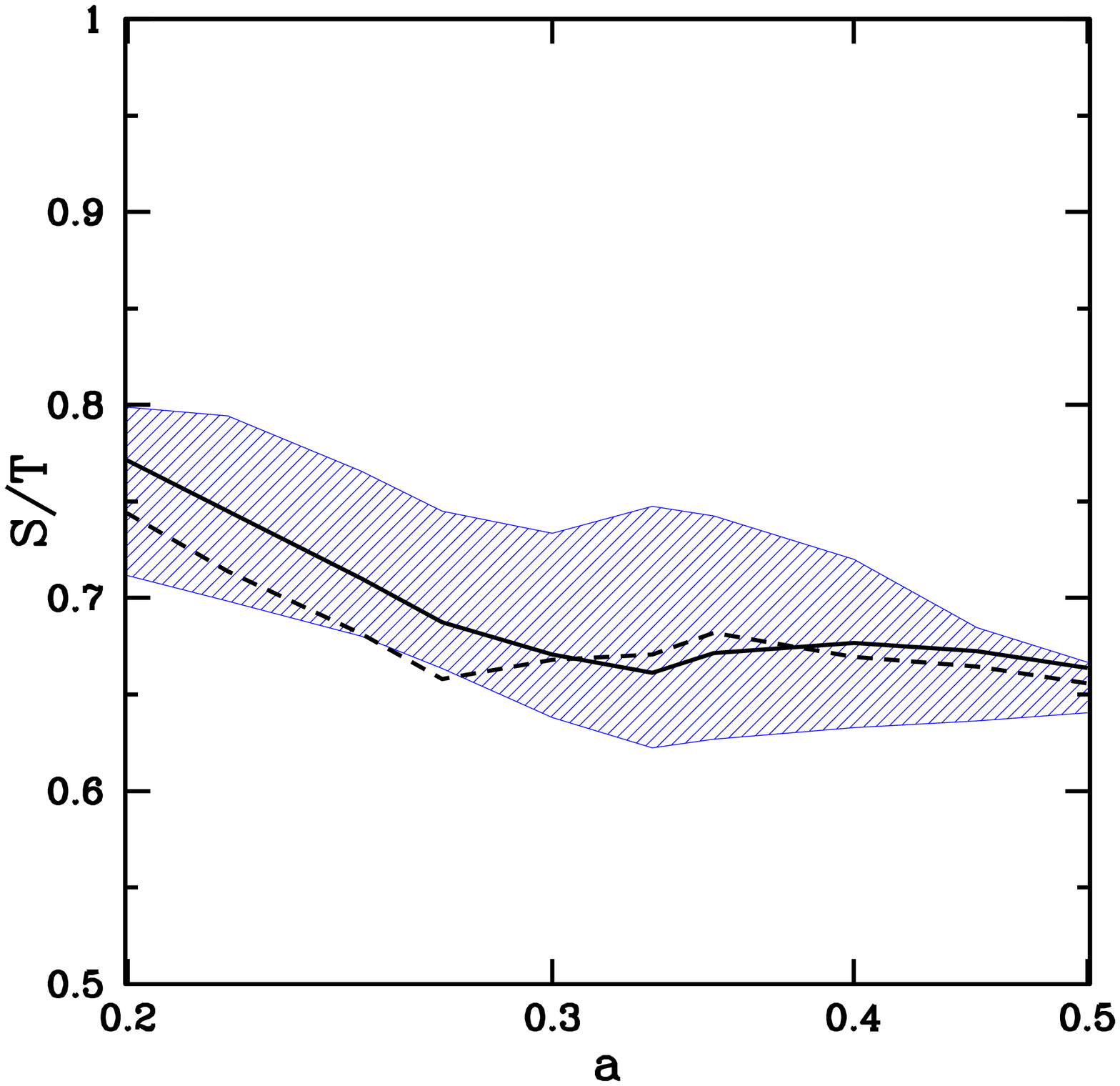}
\caption{Top: Stellar mass as a function of the Universe expansion factor for the subsample of fixed stellar mass. 
Bottom: Spheroid mass fraction for the subsample of fixed stellar mass.
The black solid lines represent the median and the blue shade regions represent the 25\% and 75\% percentiles, as in \Fig{halos}. The dashed lines are the corresponding stellar mass and spheroid fraction for the evolving sample, as shown in \Fig{Mtracks}.}
\label{fig:MfS_tight}
\end{figure}

The surface density inside a fixed radius of 1 kpc (\Fig{Sigmae1Re_Evo}) complements the information given by  $\Sigma_e$, because it does not depend explicitly on $\re$.
It gives information about the evolution of the central surface density inside 1 kpc and about the build-up of the inner spheroid.

Only the total stellar mass seems to be the relevant factor that affects $\SigmaOneKpc$.
For example, both \SFGseven \ at $z=4$ and \VLtwo \ at $z=1$ have similar masses, $\Ms \sim 10^{11} \ \msun$ and they also have similar central densities, $\SigmaOneKpc \sim 10^{10} \ \msun \kpc^{-2}$, although their effective radii are very different, $\re \sim 0.5 \kpc$ and  $\re \sim 3 \kpc$ for \SFGseven \ and \VLtwo \ respectively.
We can use the same functional form, used for the total stellar mass in \se{sample},
where $\alpha=-0.52$ and $\SigmaOneKpc(z=1)=1.7 \times 10^{10} \ \msun \kpc^{-2}$ give the best fit.
It seems that the inner mass grows as the total mass of the galaxy grows. 
This is due to the proportionality between the stellar mass inside 1 \kpc \ and the total stellar mass, because 
of the steep mass profiles and the fact that
the effective radius is always close enough to 1 \kpc. Therefore,  the mass within 1 \kpc \ is always close to $0.5 \times \Ms$.

The values of $\SigmaOneKpc \simeq 10^{10} \ \msun \kpc^{-2}$ at $z=1-2$ are similar to the observed values \citep{vDokkum14}. 
At lower redshifts, dry mergers as well as mass loss due to stellar winds from evolved populations may decrease the central density, as suggested by \citet{vDokkum14}.

\section{Redshift evolution of a subsample at fixed stellar mass}
\label{sec:fixedmass}

The previous sections was focused on the evolution, or lack of evolution, of the morphological properties as galaxies grow in mass over time at moderately high redshifts, z=4 to 1.
In this section, we study the redshift dependence of the morphological properties at a fixed stellar mass.
We create a subsample of snapshots at the available redshifts with a small range of stellar mass, $\Ms=0.5-2 \times 10^{11} \ \msun$. 
The top panel of \Fig{MfS_tight} demonstrates the constancy of the stellar mass of the selected subsample.
The variations in stellar mass between galaxies in this sample are typically less than a factor of two in a bin about $\Msun\simeq10^{11} \ \msun$.
This subsample allows us to study explicit redshift dependence.

The spheroid mass fraction of this constant-mass subsample evolves slowly with time (\Fig{MfS_tight}).
This weak evolution is very similar to the evolution of the sample of evolving galaxies.
Therefore, it seems that the spheroid mass fraction does not depend much on redshift,
as predicted by the steady-state solution of VDI at $z>1$ (DSC). 


\begin{figure}
\includegraphics[width =  0.5 \textwidth]{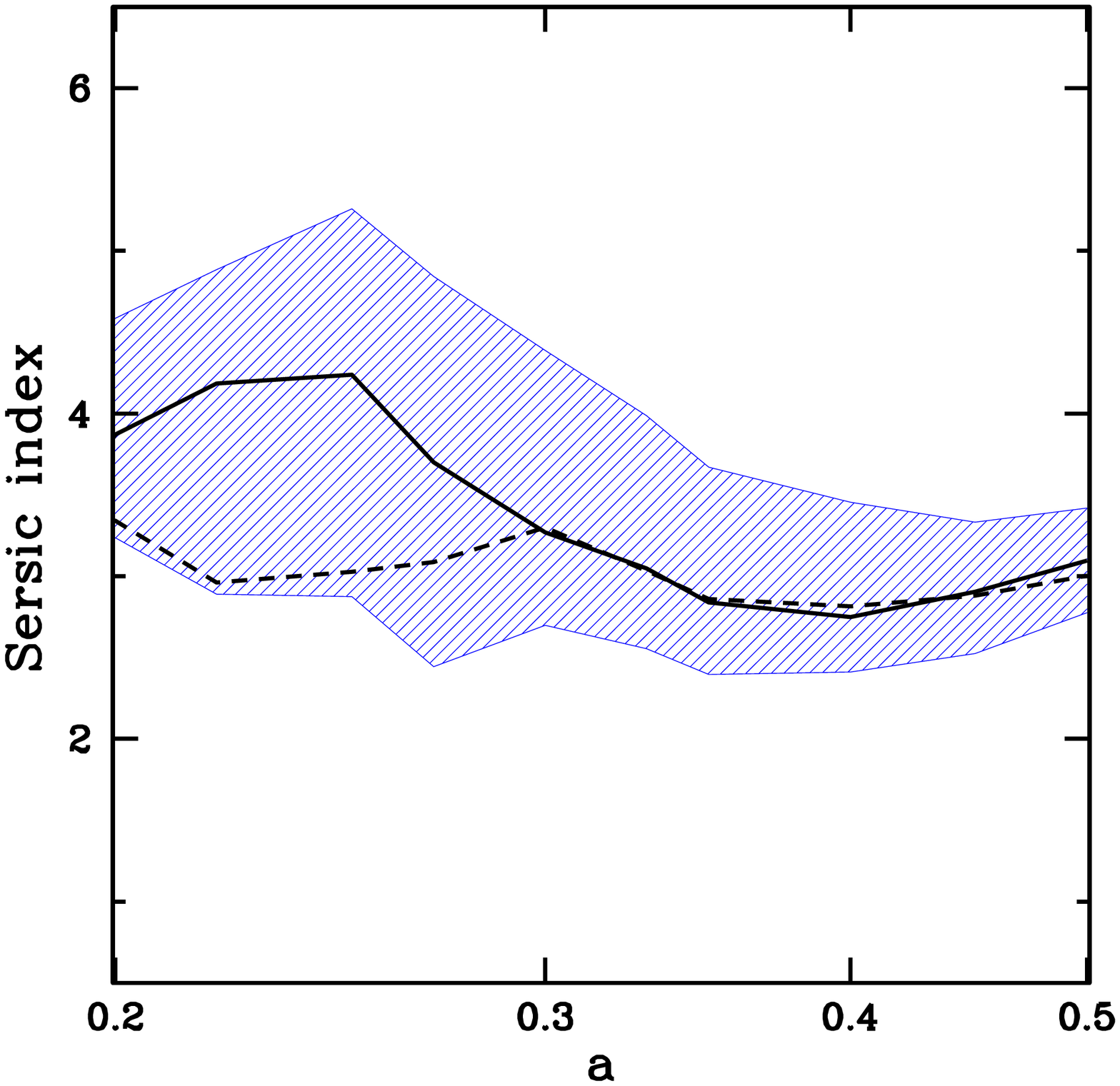}
\includegraphics[width = 0.5 \textwidth]{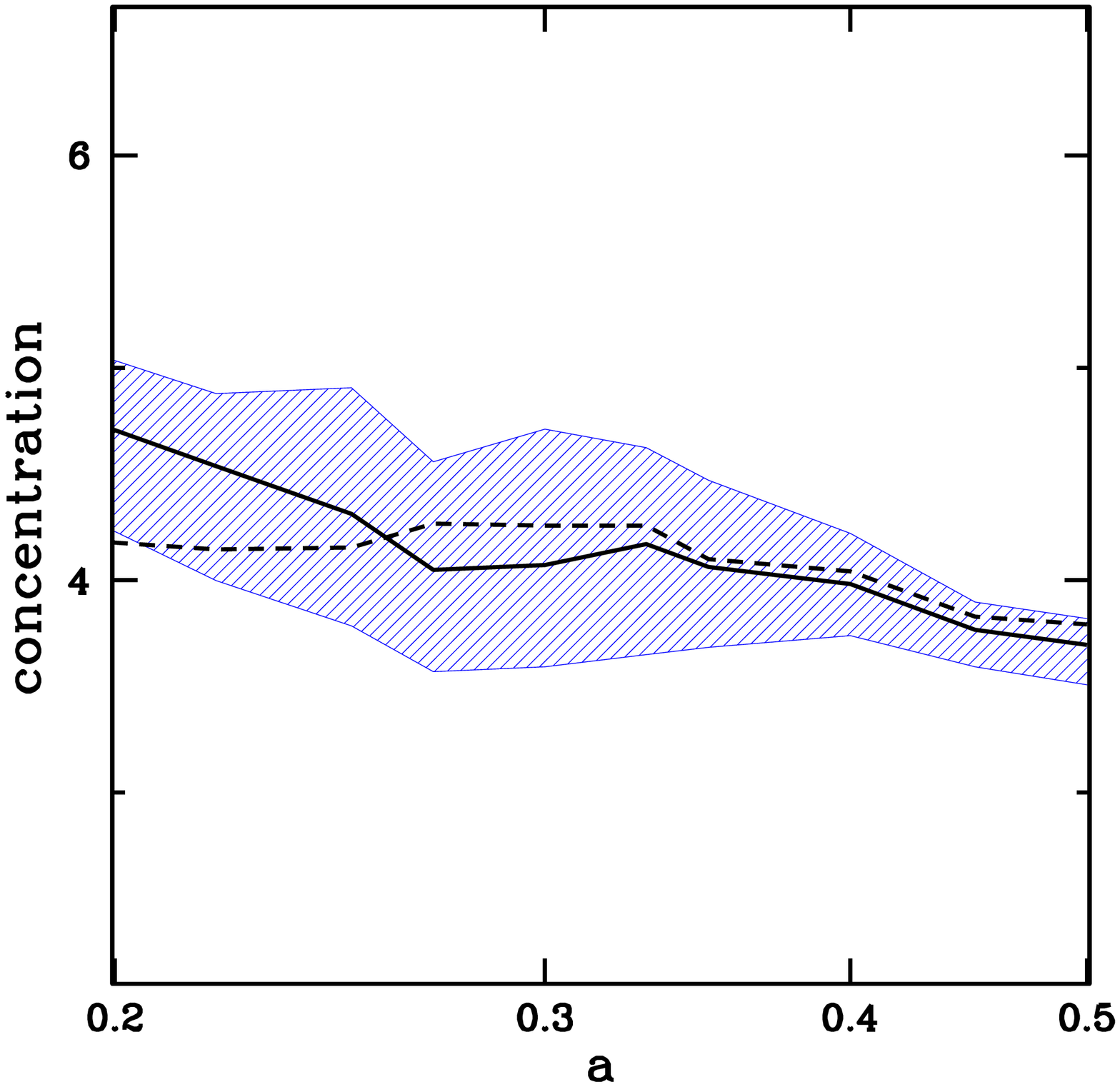}
\caption{Evolution of the sersic index (top) and concentration (bottom) for the sample of fixed stellar mass.  Lines are defined as in \Fig{MfS_tight}. }
\label{fig:n_tight}
\end{figure}

The shapes of the profiles are very similar for the subsample of fixed stellar mass (\Fig{n_tight})
 and the whole sample of growing masses.
This is expected because the spheroid fraction, which ultimately controls the shape of the total profile, has a weak redshift dependence.
However, it seems that the sersic index of the fixed-mass sample at high-z, $n(z=3-4)\sim4$, is slightly higher than the sersic index of the sample of evolving mass, $n(z=3-4)\sim3$.
Since at these redshifts the masses of the former sample are systematically higher than the masses of the latter sample, this implies that the most massive spheroids at a given redshift have the steepest profiles.
The sersic index of the spheroidal component in the fixed-mass sample exceeds $n\geq5$, while the evolving sample has spheroids with a median of $n\simeq4.5$ (\Fig{nall}). 
This could imply that the most massive galaxies at a given redshift acquired their mass earlier, when gas fractions were higher and dissipative processes were more important in making steeper density profiles.
The concentration similarly shows a weak redshift dependence. 
VDI predicts a steady-state growth of the spheroid, so this lack of explicit time dependence is not a surprise if VDI dominates the growth of the spheroids.

\begin{figure*}
\includegraphics[width = 0.99 \textwidth]{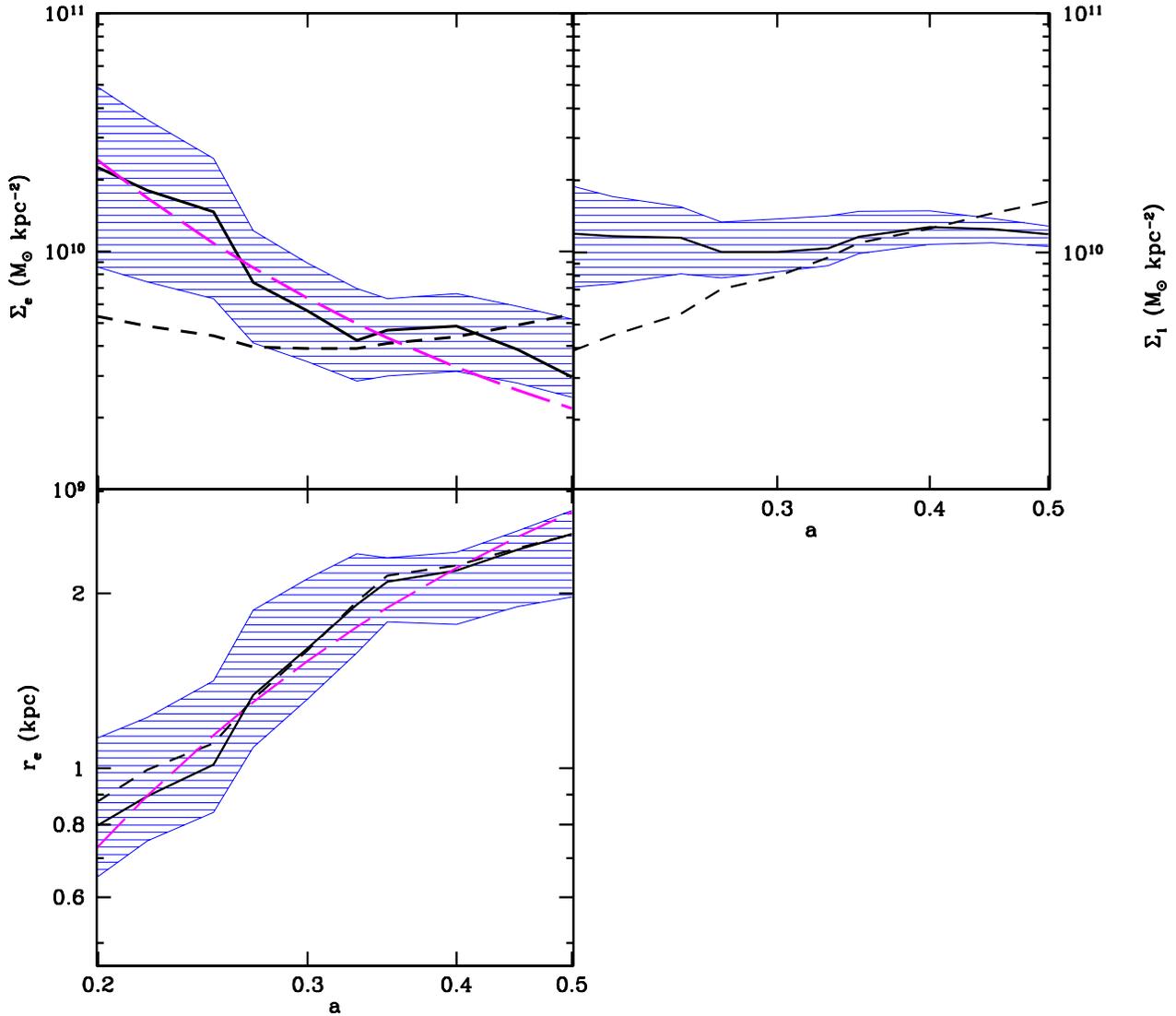}
\caption{Evolution of the efective surface density (top-left), the central surface density (top-right) and the effective radius (bottom-left) for the sample of fixed stellar mass. 
Lines are defined as in \Fig{MfS_tight} plus the dashed magenta lines, which represent the \equ{Se_full} and \equ{Re_full}.}
\label{fig:Sigmae1Re_tight}
\end{figure*}


\Fig{Sigmae1Re_tight} shows the evolution of the surface density inside the effective radius. 
It shows that galaxies of the same mass are significantly denser at higher redshifts, by almost an order of magnitude between $z=4$ and $z=1$.
This is a natural result, based on the fact that the Universe at higher redshifts was denser and all processes of mass assembly were more dissipative and more gas rich. This produces denser galaxies at higher redshifts for the same stellar mass.

At a fixed mass, $\Sigma_e \propto \re^{-2}$, so we can use the redshift dependence of $\re \propto e^{-0.4(z-1)}$, described in \S6.3. 
This gives $\Sigma_e \propto e^{0.8(z-1)}$ at a fixed mass, consistent with the strong evolution shown in \Fig{Sigmae1Re_tight}.
This redshift dependence contrasts with the fact that galaxies evolves with an approximately constant surface density (\Fig{Sigmae1Re_Evo}).
Both results can be reconciled if the mass dependence of the surface density gives the opposite redshift trend than the explicit redshift dependence. Indeed, the rate of stellar mass growth is $\alpha=-0.66$ (\se{sample}).
This gives a very small evolution of the surface density for an evolving population, $\Sigma_e \propto \Ms \re^{-2} \propto e^{0.14(z-1)}$.
Therefore, we can tentatively write the full mass and redshift dependence of the mean surface density inside the effective radius as:
\be
\Sigma_e = 2.2 \times 10^9  \msun \kpc^{-2} \left ( \frac{\Ms}{10^{11}  \msun} \right ) e^{0.8 (z-1)} 
\label{eq:Se_full}
\ee
This relation implies that more massive galaxies at a given redshift are also denser. 
For example, at $z=4$, \VLthree \ has $\Ms\sim 10^{10} \ \msun$ and $\re=1.2 \kpc$, which yields  $\Sigma_e \simeq 2 \times 10^9 \msun \kpc^{-2}$. 
In contrast, \SFGtwo \ is 10 times more massive at the same redshift but it has $\re \sim 0.9 \kpc$, which yields  $\Sigma_e \sim 3 \times 10^{10} \ \msun \kpc^{-2}$, a factor 15 times denser than \VLthree \ at the same redshift.



\Fig{Sigmae1Re_tight} also shows the evolution of the effective radius for a fixed stellar mass bin.
The evolution is very similar to the size growth of the evolving population, described earlier  (\Fig{Sigmae1Re_Evo}).
It seems that the effective radius has only a weak dependence on stellar mass, as shown in the mass-size relation at $z$=2 in \Fig{ReMs_comp}.
Ignoring this weak mass dependence, we can write the redshift dependence of the effective radius as
\be
\re = 2.7 \kpc \times e^{-0.4 (z-1)} 
\label{eq:Re_full}
\ee
This relation is a direct result of the weak evolution of the surface density for the evolving galaxies.


In section \se{Sigma1Evo}, we saw that the central surface density, the surface density inside 1 \kpc, roughly scales with  the total stellar mass. 
Then, if we take a subsample with a constant stellar mass, we expect the central surface density to be also constant and independent of redshift. This is what is shown in \Fig{Sigmae1Re_tight}.
A roughly constant stellar mass of $10^{11} \ \msun$ corresponds to a constant central surface density of $\SigmaOneKpc=10^{10} \ \msun \kpc^{-2}$.
We can tentatively write the mass dependence of the central surface density as,
\be
\SigmaOneKpc = 10^{10} \  \msun \kpc^{-2} \left ( \frac{\Ms}{10^{11}  \ \msun} \right )
\label{eq:S1_full}
\ee
The similarities between the growth of the central surface density and the growth of the stellar mass (\se{sample}) support the idea that the central surface density has a very weak explicit dependence on redshift,
after the first growth of the compact spheroid by wet contraction \citep[][Zolotov et al]{DekelBurkert}

\section{Summary and Discussion}
\label{sec:Summary}

We addressed the formation of massive and compact spheroids in the early Universe, using a set of zoom-in, AMR cosmological simulations of 17 moderately massive galaxies, $\Ms\simeq10^{11} \ \msun$ at $z\simeq1$, with a maximum resolution of 35-70 pc and a weak feedback model. 
We also include two simulations with twice better resolution, radiative feedback and lower star-formation efficiency \citep[ Zolotov et al. ]{Ceverino13}. These simulations allow us to check the robustness of the main conclusions, which 
can be summarized as follow:


\begin{itemize}
\item
Spheroid growth is driven by either  violent disc instability or mergers or by both.  At high redshifts these processes tend to be gas rich, and the dissipation naturally leads to compact spheroids.
\item
A disc/spheroid decomposition using stellar kinematics yielded a time independent spheroid fraction, $S/T\simeq0.7$. This means that the total mass is dominated by the spheroidal component, but the mass in a rotating disc is also significant.
\item
VDI-driven spheroids, much like merger-driven spheroids, have steep surface density profiles, consistent with a classical, de-Vaucouleurs profile at all times.
A sersic fitting of the combined spheroid+disc surface density profile yields a sersic index $n\simeq3$, between the high index of the spheroid ($n\simeq4$) and the low index of the disc ($n\simeq1$). 
\item
The shape of the combined profiles remains constant with time, while the spheroid and disc grow in a quasi-steady state, driven by the cosmological inflow rate and violent disk instability, as predicted by theory (DSC).
\item
After their initial growth, galaxies grow in mass and size, such that the  stellar surface density inside the effective radius evolves only little with time (\equ{Se_full}).
This is  consistent with the analytic result of galaxy growth if the halo-to-galaxy size ratio slightly declines with increasing mass and time due to angular-momentum changes and dissipative effects \citep{Dekel13}.
\item
The central surface density, defined as the stellar surface density inside a fixed radius of 1 \kpc, 
is proportional to the total mass growth, \equ{S1_full}. 
\item
Galaxies with the same stellar mass ($\Ms\simeq 10^{11} \ \msun$) at higher redshifts are denser, \equ{Se_full}, because they are smaller, \equ{Re_full}.
This is a  consequence of the fact that  the Universe at higher redshifts is denser and all processes of mass assembly  are therefore more dissipative.

\end{itemize}

%
%

The main limitation of the simulations used here is the overproduction of stars at high redshifts, when  the halos are small, $\Mv \le 10^{11} \msun$, and star formation should be very inefficient \citep{DekelSilk86}.
The star-formation efficiency is too high, and the stellar feedback is too weak. As a result, the gas
is consumed too fast, so the simulated discs have lower gas fraction than observed in discs at $z=2$, and perhaps lower SFR than observed.
A direct consequence of this high early SFR is that the stellar-to-virial mass ratio is too high in comparison to the estimates from abundance matching \citep{Moster12, Behroozi12}. 
These shortcomings compared to observations are on the order of a factor of 2, as evaluated in CDB.

The early formation of stars potentially affects the spheroid formation and properties.
First, these simulations underestimate the importance of  dissipative processes, like  VDI and wet mergers at $z\simeq1-2$. 
Second, too early star formation may overestimate the mass in the spheroid and the spheroid mass fraction.
However, the mass assembled in the spheroid at $z>3$ is only $\sim$20\% of the final mass of the spheroid at $z=1$.
Half of the mass of the spheroids at $z=1$ was actually assembled after $z=1.8.$

 Third, although all the simulated galaxies have sub- stantial
 discs of gas and stars, 
 there are no disc-dominated galaxies among our sample of massive galaxies. 
 Most probably, the thermal supernova feedback incorporated in these simulations is too weak to prevent the formation of an early spheroid, along with the gas-rich, unstable disc. Other forms of feedback, more effective in regions of high density of baryonic material, such as the center of spheroids, may efficiently eject star-forming gas and slow down star formation in the center. Radiative stellar feedback \citep{Ceverino13} or AGN feedback \citep{Bournaud11} may be good candidates for regulating star-formation in the center of clumpy, unstable discs.
However, even simulations with strong feedback and massive outflows \citep{Genel12} have unstable clumpy discs with a massive bulge at the center. This is a direct result of disc instabilities, which maintain a steady inflow of material towards the center, even in the presence of strong feedback.

%

Similar simulations with lower star-formation effi- ciencies and stronger feedback due to the addition of radiative stellar feedback  \citep{Ceverino13}  provide a better match to the observed gas fraction and stellar-to-halo mass ratio. However, the morphological properties of the massive galaxies at $z\simeq1-2$ remain qualitatively similar, 
especially for the progenitors of massive galaxies in halos of $10^{12} \Msun$ at $z\simeq1$. 
The spheroid mass fraction is reduced but
these spheroids also have classical profiles.
We thus conclude that our results concerning the classical profiles of spheroids are insensitive to variations in the strength of stellar feedback within the range favored by theoretical considerations \citep{DekelKrumholz13}.

Finally, It seems that the main effect of the too early high SFR is to cause the wet contraction of the discs \citep{DekelBurkert} to occur in most cases at very high redshifts, prior to $z\simeq4$. 
In the analyzed range $z=4-1$, after the first spheroid growth, the bulge surface density is already high, but the quenching of star formation is not yet complete because of the relatively weak feedback.
Indeed, in the simulations with radiative feedback we see more extended discs at $z\simeq4-2$, with the disc shrinkage  occurring later, first to star-forming compact systems (``blue nuggets"), followed by quenching to passive compact spheroids (``red nuggets"). This scenario, as predicted in \citet{DekelBurkert}, is analyzed using the simulations with radiative feedback in Zolotov et al. (in preparation). The classical-spheroid morphology is similar to our findings here.


\section*{Acknowledgments} 
 
We acknowledge stimulating discussions with Sandra Faber, David Koo, Andreas Burkert, Shardha Jogee and Guillermo Barro.
The simulations were performed in the astro cluster at HU, 
at the National Energy Research Scientific Computing Center (NERSC),  
Lawrence Berkeley National Laboratory, 
at NASA Advanced Supercomputing (NAS) at NASA Ames Research Center.
This work was partly supported, by MINECO grant AYA2012-31101,
by MICINN grant  AYA-2009-13875-C03-02, 
by ISF grant 24/12,
by GIF grant G-1052-104.7/2009,
by a DIP grant,
by NSF grant AST-1010033,
and by the I-CORE Program of the PBC and The ISF grant 1829/12.
DC is a Juan de la Cierva fellow.

\bibliographystyle{mn2e}
\bibliography{morphov7}

\begin{thebibliography}{}

\bibitem[\protect\citeauthoryear{{Agertz}, {Teyssier} \& {Moore}}{{Agertz}
  et~al.}{2009}]{Agertz09b}
{Agertz} O.,  {Teyssier} R.,    {Moore} B.,  2009, \mnras, 397, L64

\bibitem[\protect\citeauthoryear{{Barro}, {Faber}, {P{\'e}rez-Gonz{\'a}lez},
  {Koo}, {Williams}, {Kocevski} \& {et al.,}}{{Barro} et~al.}{2013}]{Barro13}
{Barro} G.,  {Faber} S.~M.,  {P{\'e}rez-Gonz{\'a}lez} P.~G.,  {Koo} D.~C.,
  {Williams} C.~C.,  {Kocevski} D.~D.,    {et al.,} 2013, \apj, 765, 104

\bibitem[\protect\citeauthoryear{{Barro}, {Faber}, {Perez-Gonzalez},
  {Pacifici}, {Trump} \& {et al.,}}{{Barro} et~al.}{2014}]{Barro14}
{Barro} G.,  {Faber} S.~M.,  {Perez-Gonzalez} P.~G.,  {Pacifici} C.,  {Trump}
  J.~R.,    {et al.,} \ 2014, \apj, 791, 52 

\bibitem[\protect\citeauthoryear{{Barro}, {Trump}, {Koo}, {Dekel}, {Kassin},
  {Kocevski}, {Faber}, {van der Wel} \& {et al.,}}{{Barro}
  et~al.}{2014b}]{Barro14b}
{Barro} G.,  {Trump} J.~R.,  {Koo} D.~C.,  {Dekel} A.,  {Kassin} S.~A.,
  {Kocevski} D.~D.,  {Faber} S.~M.,  {van der Wel} A.,    {et al.,} \ 2014, \apj, 795, 145

\bibitem[\protect\citeauthoryear{{Behroozi}, {Wechsler} \& {Conroy}}{{Behroozi}
  et~al.}{2013}]{Behroozi12}
{Behroozi} P.~S.,  {Wechsler} R.~H.,    {Conroy} C.,  2013, \apj, 770, 57

\bibitem[\protect\citeauthoryear{{Birnboim} \& {Dekel}}{{Birnboim} \&
  {Dekel}}{2003}]{bd03}
{Birnboim} Y.,  {Dekel} A.,  2003, \mnras, 345, 349

\bibitem[\protect\citeauthoryear{{Bournaud}, {Dekel}, {Teyssier}, {Cacciato},
  {Daddi}, {Juneau} \& {Shankar}}{{Bournaud} et~al.}{2011}]{Bournaud11}
{Bournaud} F.,  {Dekel} A.,  {Teyssier} R.,  {Cacciato} M.,  {Daddi} E.,
  {Juneau} S.,    {Shankar} F.,  2011, \apjl, 741, L33

\bibitem[\protect\citeauthoryear{{Bournaud} \& {Elmegreen}}{{Bournaud} \&
  {Elmegreen}}{2009}]{Bournaud09}
{Bournaud} F.,  {Elmegreen} B.~G.,  2009, \apjl, 694, L158

\bibitem[\protect\citeauthoryear{{Bournaud}, {Elmegreen} \&
  {Elmegreen}}{{Bournaud} et~al.}{2007}]{Bournaud07}
{Bournaud} F.,  {Elmegreen} B.~G.,    {Elmegreen} D.~M.,  2007, \apj, 670, 237

\bibitem[\protect\citeauthoryear{{Buitrago}, {Trujillo}, {Conselice},
  {Bouwens}, {Dickinson} \& {Yan}}{{Buitrago} et~al.}{2008}]{Buitrago08}
{Buitrago} F.,  {Trujillo} I.,  {Conselice} C.~J.,  {Bouwens} R.~J.,
  {Dickinson} M.,    {Yan} H.,  2008, \apjl, 687, L61

\bibitem[\protect\citeauthoryear{{Bullock}, {Dekel}, {Kolatt}, {Kravtsov},
  {Klypin}, {Porciani} \& {Primack}}{{Bullock} et~al.}{2001}]{Bullock01}
{Bullock} J.~S.,  {Dekel} A.,  {Kolatt} T.~S.,  {Kravtsov} A.~V.,  {Klypin}
  A.~A.,  {Porciani} C.,    {Primack} J.~R.,  2001, \apj, 555, 240

\bibitem[\protect\citeauthoryear{{Cacciato}, {Dekel} \& {Genel}}{{Cacciato}
  et~al.}{2012}]{Cacciato12}
{Cacciato} M.,  {Dekel} A.,    {Genel} S.,  2012, \mnras, 421, 818

\bibitem[\protect\citeauthoryear{{Cassata}, {Giavalisco}, {Guo}, {Ferguson},
  {Koekemoer}, {Renzini}, {Fontana} \& {et al.,}}{{Cassata}
  et~al.}{2010}]{Cassata10}
{Cassata} P.,  {Giavalisco} M.,  {Guo} Y.,  {Ferguson} H.,  {Koekemoer} A.~M.,
  {Renzini} A.,  {Fontana} A.,    {et al.,} 2010, \apjl, 714, L79

\bibitem[\protect\citeauthoryear{{Ceverino}, {Dekel} \& {Bournaud}}{{Ceverino}
  et~al.}{2010}]{CDB}
{Ceverino} D.,  {Dekel} A.,    {Bournaud} F.,  2010, \mnras, 404, 2151

\bibitem[\protect\citeauthoryear{{Ceverino}, {Dekel}, {Mandelker}, {Bournaud},
  {Burkert}, {Genzel} \& {Primack}}{{Ceverino} et~al.}{2012}]{Ceverino12}
{Ceverino} D.,  {Dekel} A.,  {Mandelker} N.,  {Bournaud} F.,  {Burkert} A.,
  {Genzel} R.,    {Primack} J.,  2012, \mnras, 420, 3490

\bibitem[\protect\citeauthoryear{{Ceverino} \& {Klypin}}{{Ceverino} \&
  {Klypin}}{2009}]{Ceverino09}
{Ceverino} D.,  {Klypin} A.,  2009, \apj, 695, 292

\bibitem[\protect\citeauthoryear{{Ceverino}, {Klypin}, {Klimek},
  {Trujillo-Gomez}, {Churchill}, {Primack} \& {Dekel}}{{Ceverino}
  et~al.}{2014}]{Ceverino13}
{Ceverino} D.,  {Klypin} A.,  {Klimek} E.,  {Trujillo-Gomez} S.,  {Churchill}
  C.~W.,  {Primack} J.,    {Dekel} A.,  \ 2014, \mnras, 442, 1545

\bibitem[\protect\citeauthoryear{{Cimatti}, {Cassata}, {Pozzetti}, {Kurk},
  {Mignoli}, {Renzini}, {Daddi}, {Bolzonella}, {Brusa}, {Rodighiero},
  {Dickinson}, {Franceschini}, {Zamorani}, {Berta}, {Rosati} \&
  {Halliday}}{{Cimatti} et~al.}{2008}]{Cimatti08}
{Cimatti} A.,  {Cassata} P.,  {Pozzetti} L.,  {Kurk} J.,  {Mignoli} M.,
  {Renzini} A.,  {Daddi} E.,  {Bolzonella} M.,  {Brusa} M.,  {Rodighiero} G.,
  {Dickinson} M.,  {Franceschini} A.,  {Zamorani} G.,  {Berta} S.,  {Rosati}
  P.,    {Halliday} C.,  2008, \aap, 482, 21

\bibitem[\protect\citeauthoryear{{Conselice}, {Bluck}, {Buitrago}, {Bauer},
  {Gr{\"u}tzbauch}, {Bouwens}, {Bevan}, {Mortlock} \& {et al.,}}{{Conselice}
  et~al.}{2011}]{Conselice11}
{Conselice} C.~J.,  {Bluck} A.~F.~L.,  {Buitrago} F.,  {Bauer} A.~E.,
  {Gr{\"u}tzbauch} R.,  {Bouwens} R.~J.,  {Bevan} S.,  {Mortlock} A.,    {et
  al.,} 2011, \mnras, 413, 80

\bibitem[\protect\citeauthoryear{{Cowie}, {Hu} \& {Songaila}}{{Cowie}
  et~al.}{1995}]{Cowie95}
{Cowie} L.~L.,  {Hu} E.~M.,    {Songaila} A.,  1995, \aj, 110, 1576

\bibitem[\protect\citeauthoryear{{Cox}, {Dutta}, {Di Matteo}, {Hernquist},
  {Hopkins}, {Robertson} \& {Springel}}{{Cox} et~al.}{2006}]{Cox06}
{Cox} T.~J.,  {Dutta} S.~N.,  {Di Matteo} T.,  {Hernquist} L.,  {Hopkins}
  P.~F.,  {Robertson} B.,    {Springel} V.,  2006, \apj, 650, 791

\bibitem[\protect\citeauthoryear{{Daddi}, {Renzini}, {Pirzkal}, {Cimatti},
  {Malhotra}, {Stiavelli}, {Xu}, {Pasquali}, {Rhoads}, {Brusa}, {di Serego
  Alighieri}, {Ferguson}, {Koekemoer}, {Moustakas}, {Panagia} \&
  {Windhorst}}{{Daddi} et~al.}{2005}]{Daddi05}
{Daddi} E.,  {Renzini} A.,  {Pirzkal} N.,  {Cimatti} A.,  {Malhotra} S.,
  {Stiavelli} M.,  {Xu} C.,  {Pasquali} A.,  {Rhoads} J.~E.,  {Brusa} M.,  {di
  Serego Alighieri} S.,  {Ferguson} H.~C.,  {Koekemoer} A.~M.,  {Moustakas}
  L.~A.,  {Panagia} N.,    {Windhorst} R.~A.,  2005, \apj, 626, 680

\bibitem[\protect\citeauthoryear{{Danovich}, {Dekel}, {Hahn}, {Ceverino} \&
  {Primack}}{{Danovich} et~al.}{2014}]{Danovich14}
{Danovich} M.,  {Dekel} A.,  {Hahn} O.,  {Ceverino} D.,    {Primack} J.,  2014,
  arXiv:1407.7129

\bibitem[\protect\citeauthoryear{{de Vaucouleurs}}{{de
  Vaucouleurs}}{1948}]{Vaucouleurs}
{de Vaucouleurs} G.,  1948, Annales d'Astrophysique, 11, 247

\bibitem[\protect\citeauthoryear{{Dekel} \& {Birnboim}}{{Dekel} \&
  {Birnboim}}{2006}]{db06}
{Dekel} A.,  {Birnboim} Y.,  2006, \mnras, 368, 2

\bibitem[\protect\citeauthoryear{{Dekel}, {Birnboim}, {Engel}, {Freundlich},
  {Goerdt}, {Mumcuoglu}, {Neistein}, {Pichon}, {Teyssier} \& {Zinger}}{{Dekel}
  et~al.}{2009}]{dekel09}
{Dekel} A.,  {Birnboim} Y.,  {Engel} G.,  {Freundlich} J.,  {Goerdt} T.,
  {Mumcuoglu} M.,  {Neistein} E.,  {Pichon} C.,  {Teyssier} R.,    {Zinger} E.,
   2009, \nat, 457, 451

\bibitem[\protect\citeauthoryear{{Dekel} \& {Burkert}}{{Dekel} \&
  {Burkert}}{2014}]{DekelBurkert}
{Dekel} A.,  {Burkert} A.,  2014, \mnras, 438, 1870

\bibitem[\protect\citeauthoryear{{Dekel} \& {Cox}}{{Dekel} \&
  {Cox}}{2006}]{DekelCox}
{Dekel} A.,  {Cox} T.~J.,  2006, \mnras, 370, 1445

\bibitem[\protect\citeauthoryear{{Dekel} \& {Krumholz}}{{Dekel} \&
  {Krumholz}}{2013}]{DekelKrumholz13}
{Dekel} A.,  {Krumholz} M.~R.,  2013, \mnras, 432, 455

\bibitem[\protect\citeauthoryear{{Dekel}, {Sari} \& {Ceverino}}{{Dekel}
  et~al.}{2009}]{DSC}
{Dekel} A.,  {Sari} R.,    {Ceverino} D.,  2009, \apj, 703, 785

\bibitem[\protect\citeauthoryear{{Dekel} \& {Silk}}{{Dekel} \&
  {Silk}}{1986}]{DekelSilk86}
{Dekel} A.,  {Silk} J.,  1986, \apj, 303, 39

\bibitem[\protect\citeauthoryear{{Dekel}, {Zolotov}, {Tweed}, {Cacciato},
  {Ceverino} \& {Primack}}{{Dekel} et~al.}{2013}]{Dekel13}
{Dekel} A.,  {Zolotov} A.,  {Tweed} D.,  {Cacciato} M.,  {Ceverino} D.,
  {Primack} J.~R.,  2013, \mnras, 435, 999

\bibitem[\protect\citeauthoryear{{Dickinson}, {Papovich}, {Ferguson} \&
  {Budav{\'a}ri}}{{Dickinson} et~al.}{2003}]{Dickinson03}
{Dickinson} M.,  {Papovich} C.,  {Ferguson} H.~C.,    {Budav{\'a}ri} T.,  2003,
  \apj, 587, 25

\bibitem[\protect\citeauthoryear{{Elmegreen}, {Bournaud} \&
  {Elmegreen}}{{Elmegreen} et~al.}{2008}]{Elmegreen08a}
{Elmegreen} B.~G.,  {Bournaud} F.,    {Elmegreen} D.~M.,  2008, \apj, 688, 67

\bibitem[\protect\citeauthoryear{{Elmegreen} \& {Elmegreen}}{{Elmegreen} \&
  {Elmegreen}}{2006}]{Elmegreen06}
{Elmegreen} B.~G.,  {Elmegreen} D.~M.,  2006, \apj, 650, 644

\bibitem[\protect\citeauthoryear{{Elmegreen}, {Elmegreen}, {Fernandez} \&
  {Lemonias}}{{Elmegreen} et~al.}{2009}]{Elmegreen09}
{Elmegreen} B.~G.,  {Elmegreen} D.~M.,  {Fernandez} M.~X.,    {Lemonias} J.~J.,
   2009, \apj, 692, 12

\bibitem[\protect\citeauthoryear{{Elmegreen}, {Elmegreen} \&
  {Hirst}}{{Elmegreen} et~al.}{2004}]{Elmegreen04b}
{Elmegreen} D.~M.,  {Elmegreen} B.~G.,    {Hirst} A.~C.,  2004, \apjl, 604, L21

\bibitem[\protect\citeauthoryear{{Elmegreen}, {Elmegreen}, {Ravindranath} \&
  {Coe}}{{Elmegreen} et~al.}{2007}]{Elmegreen07}
{Elmegreen} D.~M.,  {Elmegreen} B.~G.,  {Ravindranath} S.,    {Coe} D.~A.,
  2007, \apj, 658, 763

\bibitem[\protect\citeauthoryear{{Elmegreen}, {Elmegreen}, {Rubin} \&
  {Schaffer}}{{Elmegreen} et~al.}{2005}]{Elmegreen05}
{Elmegreen} D.~M.,  {Elmegreen} B.~G.,  {Rubin} D.~S.,    {Schaffer} M.~A.,
  2005, \apj, 631, 85

\bibitem[\protect\citeauthoryear{{Fall} \& {Efstathiou}}{{Fall} \&
  {Efstathiou}}{1980}]{FallEfstathiou80}
{Fall} S.~M.,  {Efstathiou} G.,  1980, \mnras, 193, 189

\bibitem[\protect\citeauthoryear{{Ferland}, {Korista}, {Verner}, {Ferguson},
  {Kingdon} \& {Verner}}{{Ferland} et~al.}{1998}]{Ferland98}
{Ferland} G.~J.,  {Korista} K.~T.,  {Verner} D.~A.,  {Ferguson} J.~W.,
  {Kingdon} J.~B.,    {Verner} E.~M.,  1998, \pasp, 110, 761

\bibitem[\protect\citeauthoryear{{Fisher} \& {Drory}}{{Fisher} \&
  {Drory}}{2008}]{FisherDrory08}
{Fisher} D.~B.,  {Drory} N.,  2008, \aj, 136, 773

\bibitem[\protect\citeauthoryear{{Forbes}, {Krumholz} \& {Burkert}}{{Forbes}
  et~al.}{2012}]{Forbes12}
{Forbes} J.,  {Krumholz} M.,    {Burkert} A.,  2012, \apj, 754, 48

\bibitem[\protect\citeauthoryear{{Forbes}, {Krumholz}, {Burkert} \&
  {Dekel}}{{Forbes} et~al.}{2014}]{Forbes14}
{Forbes} J.~C.,  {Krumholz} M.~R.,  {Burkert} A.,    {Dekel} A.,  2014, \mnras,
  438, 1552

\bibitem[\protect\citeauthoryear{{F{\"o}rster Schreiber}, {Genzel},
  {Bouch{\'e}}, {Cresci}, {Davies}, {Buschkamp}, {Shapiro} \& {et
  al.,}}{{F{\"o}rster Schreiber} et~al.}{2009}]{Forster09}
{F{\"o}rster Schreiber} N.~M.,  {Genzel} R.,  {Bouch{\'e}} N.,  {Cresci} G.,
  {Davies} R.,  {Buschkamp} P.,  {Shapiro} K.,    {et al.,} 2009, \apj, 706,
  1364

\bibitem[\protect\citeauthoryear{{F{\"o}rster Schreiber}, {Shapley}, {Erb},
  {Genzel}, {Steidel}, {Bouch{\'e}}, {Cresci} \& {Davies}}{{F{\"o}rster
  Schreiber} et~al.}{2011}]{Forster11a}
{F{\"o}rster Schreiber} N.~M.,  {Shapley} A.~E.,  {Erb} D.~K.,  {Genzel} R.,
  {Steidel} C.~C.,  {Bouch{\'e}} N.,  {Cresci} G.,    {Davies} R.,  2011, \apj,
  731, 65

\bibitem[\protect\citeauthoryear{{Gadotti}}{{Gadotti}}{2009}]{Gadotti09}
{Gadotti} D.~A.,  2009, \mnras, 393, 1531

\bibitem[\protect\citeauthoryear{{Gammie}}{{Gammie}}{2001}]{Gammie01}
{Gammie} C.~F.,  2001, \apj, 553, 174

\bibitem[\protect\citeauthoryear{{Genel}, {Genzel}, {Bouch{\'e}}, {Sternberg},
  {Naab} \& {et al.,}}{{Genel} et~al.}{2008}]{Genel08}
{Genel} S.,  {Genzel} R.,  {Bouch{\'e}} N.,  {Sternberg} A.,  {Naab} T.,    {et
  al.,} 2008, \apj, 688, 789

\bibitem[\protect\citeauthoryear{{Genel}, {Naab}, {Genzel}, {F{\"o}rster
  Schreiber}, {Sternberg}, {Oser}, {Johansson}, {Dav{\'e}}, {Oppenheimer} \&
  {Burkert}}{{Genel} et~al.}{2012}]{Genel12}
{Genel} S.,  {Naab} T.,  {Genzel} R.,  {F{\"o}rster Schreiber} N.~M.,
  {Sternberg} A.,  {Oser} L.,  {Johansson} P.~H.,  {Dav{\'e}} R.,
  {Oppenheimer} B.~D.,    {Burkert} A.,  2012, \apj, 745, 11

\bibitem[\protect\citeauthoryear{{Genzel}, {Burkert}, {Bouch{\'e}}, {Cresci},
  {F{\"o}rster Schreiber}, {Shapley}, {Shapiro} \& {et al.,}}{{Genzel}
  et~al.}{2008}]{Genzel08}
{Genzel} R.,  {Burkert} A.,  {Bouch{\'e}} N.,  {Cresci} G.,  {F{\"o}rster
  Schreiber} N.~M.,  {Shapley} A.,  {Shapiro} K.,    {et al.,} 2008, \apj, 687,
  59

\bibitem[\protect\citeauthoryear{{Genzel}, {F{\"o}rster Schreiber}, {Lang},
  {Tacchella}, {Tacconi}, {Wuyts}, {Bandara}, {Burkert} \& {et al.,}}{{Genzel}
  et~al.}{2014}]{Genzel14}
{Genzel} R.,  {F{\"o}rster Schreiber} N.~M.,  {Lang} P.,  {Tacchella} S.,
  {Tacconi} L.~J.,  {Wuyts} S.,  {Bandara} K.,  {Burkert} A.,    {et al.,}
  2014, \apj, 785, 75

\bibitem[\protect\citeauthoryear{{Genzel}, {Tacconi}, {Eisenhauer},
  {F{\"o}rster Schreiber}, {Cimatti}, {Daddi}, {Bouch{\'e}} \& {et
  al.,}}{{Genzel} et~al.}{2006}]{Genzel06}
{Genzel} R.,  {Tacconi} L.~J.,  {Eisenhauer} F.,  {F{\"o}rster Schreiber}
  N.~M.,  {Cimatti} A.,  {Daddi} E.,  {Bouch{\'e}} N.,    {et al.,} 2006, \nat,
  442, 786

\bibitem[\protect\citeauthoryear{{Governato}, {Brook}, {Mayer}, {Brooks},
  {Rhee}, {Wadsley}, {Jonsson}, {Willman}, {Stinson}, {Quinn} \&
  {Madau}}{{Governato} et~al.}{2010}]{Governato10}
{Governato} F.,  {Brook} C.,  {Mayer} L.,  {Brooks} A.,  {Rhee} G.,  {Wadsley}
  J.,  {Jonsson} P.,  {Willman} B.,  {Stinson} G.,  {Quinn} T.,    {Madau} P.,
  2010, \nat, 463, 203

\bibitem[\protect\citeauthoryear{{Guedes}, {Callegari}, {Madau} \&
  {Mayer}}{{Guedes} et~al.}{2011}]{Guedes11}
{Guedes} J.,  {Callegari} S.,  {Madau} P.,    {Mayer} L.,  2011, \apj, 742, 76

\bibitem[\protect\citeauthoryear{{Guo}, {Ferguson}, {Bell}  \& {et al.,}}{{Guo} et~al.}{2014}]{Guo14}
{Guo}, Y., {Ferguson}, H.~C., 
{Bell}, E.~F., et al.\ 2014, arXiv:1410.7398

\bibitem[\protect\citeauthoryear{{Guo}, {Giavalisco}, {Ferguson},
  {Cassata}, \& {Koekemoer}}{{Guo} et~al.}{2012}]{Guo12}
{Guo}, Y., {Giavalisco}, M., 
{Ferguson}, H.~C., {Cassata}, P., \& {Koekemoer}, A.~M.\ 2012, \apj, 757, 120 

\bibitem[\protect\citeauthoryear{{Haardt} \& {Madau}}{{Haardt} \&
  {Madau}}{1996}]{HaardtMadau96}
{Haardt} F.,  {Madau} P.,  1996, \apj, 461, 20

\bibitem[\protect\citeauthoryear{{Hopkins}, {Hernquist}, {Cox}, {Di Matteo},
  {Robertson} \& {Springel}}{{Hopkins} et~al.}{2006}]{Hopkins06}
{Hopkins} P.~F.,  {Hernquist} L.,  {Cox} T.~J.,  {Di Matteo} T.,  {Robertson}
  B.,    {Springel} V.,  2006, \apjs, 163, 1

\bibitem[\protect\citeauthoryear{{Immeli}, {Samland}, {Westera} \&
  {Gerhard}}{{Immeli} et~al.}{2004}]{Immeli04}
{Immeli} A.,  {Samland} M.,  {Westera} P.,    {Gerhard} O.,  2004, \apj, 611,
  20

\bibitem[\protect\citeauthoryear{{Inoue} \& {Saitoh}}{{Inoue} \&
  {Saitoh}}{2012}]{inoue12}
{Inoue} S.,  {Saitoh} T.~R.,  2012, \mnras, 422, 1902

\bibitem[\protect\citeauthoryear{{Kennicutt}
  Jr.}{{Kennicutt}}{1998}]{Kennicutt98}
{Kennicutt} Jr. R.~C.,  1998, \apj, 498, 541

\bibitem[\protect\citeauthoryear{{Kere{\v s}}, {Katz}, {Weinberg} \&
  {Dav{\'e}}}{{Kere{\v s}} et~al.}{2005}]{keres05}
{Kere{\v s}} D.,  {Katz} N.,  {Weinberg} D.~H.,    {Dav{\'e}} R.,  2005,
  \mnras, 363, 2

\bibitem[\protect\citeauthoryear{{Komatsu}, {Dunkley}, {Nolta}, {Bennett},
  {Gold}, {Hinshaw}, {Jarosik}, {Larson}, {Limon}, {Page}, {Spergel},
  {Halpern}, {Hill}, {Kogut}, {Meyer}, {Tucker}, {Weiland}, {Wollack} \&
  {Wright}}{{Komatsu} et~al.}{2009}]{WMAP5}
{Komatsu} E.,  {Dunkley} J.,  {Nolta} M.~R.,  {Bennett} C.~L.,  {Gold} B.,
  {Hinshaw} G.,  {Jarosik} N.,  {Larson} D.,  {Limon} M.,  {Page} L.,
  {Spergel} D.~N.,  {Halpern} M.,  {Hill} R.~S.,  {Kogut} A.,  {Meyer} S.~S.,
  {Tucker} G.~S.,  {Weiland} J.~L.,  {Wollack} E.,    {Wright} E.~L.,  2009,
  \apjs, 180, 330

\bibitem[\protect\citeauthoryear{{Kormendy} \& {Kennicutt} Jr.}{{Kormendy} \&
  {Kennicutt}}{2004}]{Kormendy04}
{Kormendy} J.,  {Kennicutt} Jr. R.~C.,  2004, \araa, 42, 603

\bibitem[\protect\citeauthoryear{{Kravtsov}, {Vikhlinin} \&
  {Meshscheryakov}}{{Kravtsov} et~al.}{2014}]{Kravtsov14}
{Kravtsov} A.,  {Vikhlinin} A.,    {Meshscheryakov} A.,  2014, arXiv:1401.7329

\bibitem[\protect\citeauthoryear{{Kravtsov}}{{Kravtsov}}{2003}]{Kravtsov03}
{Kravtsov} A.~V.,  2003, \apjl, 590, L1

\bibitem[\protect\citeauthoryear{{Kravtsov}, {Klypin} \& {Khokhlov}}{{Kravtsov}
  et~al.}{1997}]{Kravtsov97}
{Kravtsov} A.~V.,  {Klypin} A.~A.,    {Khokhlov} A.~M.,  1997, \apjs, 111, 73

\bibitem[\protect\citeauthoryear{{Kriek}, {van Dokkum}, {Franx}, {Illingworth}
  \& {Magee}}{{Kriek} et~al.}{2009}]{Kriek09a}
{Kriek} M.,  {van Dokkum} P.~G.,  {Franx} M.,  {Illingworth} G.~D.,    {Magee}
  D.~K.,  2009, \apjl, 705, L71

\bibitem[\protect\citeauthoryear{{Kriek}, {van Dokkum}, {Labb{\'e}}, {Franx},
  {Illingworth}, {Marchesini} \& {Quadri}}{{Kriek} et~al.}{2009}]{Kriek09b}
{Kriek} M.,  {van Dokkum} P.~G.,  {Labb{\'e}} I.,  {Franx} M.,  {Illingworth}
  G.~D.,  {Marchesini} D.,    {Quadri} R.~F.,  2009, \apj, 700, 221

\bibitem[\protect\citeauthoryear{{Krogager}, {Zirm}, {Toft}, {Man} \&
  {Brammer}}{{Krogager} et~al.}{2013}]{Krogager13}
{Krogager} J.-K.,  {Zirm} A.~W.,  {Toft} S.,  {Man} A.,    {Brammer} G.,  2013,
  arXiv:1309.6316 

\bibitem[\protect\citeauthoryear{{Krumholz} \& {Burkert}}{{Krumholz} \&
  {Burkert}}{2010}]{krumholz_burkert10}
{Krumholz} M.,  {Burkert} A.,  2010, \apj, 724, 895

\bibitem[\protect\citeauthoryear{{Laurikainen}, {Salo}, {Buta} \&
  {Knapen}}{{Laurikainen} et~al.}{2007}]{Laurikainen07}
{Laurikainen} E.,  {Salo} H.,  {Buta} R.,    {Knapen} J.~H.,  2007, \mnras,
  381, 401

\bibitem[\protect\citeauthoryear{{Law}, {Steidel}, {Erb}, {Larkin}, {Pettini},
  {Shapley} \& {Wright}}{{Law} et~al.}{2009}]{Law09}
{Law} D.~R.,  {Steidel} C.~C.,  {Erb} D.~K.,  {Larkin} J.~E.,  {Pettini} M.,
  {Shapley} A.~E.,    {Wright} S.~A.,  2009, \apj, 697, 2057

\bibitem[\protect\citeauthoryear{{Law}, {Steidel}, {Shapley}, {Nagy}, {Reddy}
  \& {Erb}}{{Law} et~al.}{2012}]{Law12}
{Law} D.~R.,  {Steidel} C.~C.,  {Shapley} A.~E.,  {Nagy} S.~R.,  {Reddy} N.~A.,
     {Erb} D.~K.,  2012, \apj, 745, 85

\bibitem[\protect\citeauthoryear{{Madau} \& {Dickinson}}{{Madau} \&
  {Dickinson}}{2014}]{Madau14}
{Madau} P.,  {Dickinson} M.,  2014, \ 2014, \araa, 52, 415 

\bibitem[\protect\citeauthoryear{{Mandelker}, {Dekel}, {Ceverino}, {Tweed},
  {Moody} \& {Primack}}{{Mandelker} et~al.}{2014}]{Mandelker}
{Mandelker} N.,  {Dekel} A.,  {Ceverino} D.,  {Tweed} D.,  {Moody} C.~E.,
  {Primack} J.,  2014, \mnras, 443, 3675

\bibitem[\protect\citeauthoryear{{Mo}, {Mao} \& {White}}{{Mo}
  et~al.}{1998}]{MoMaoWhite98}
{Mo} H.~J.,  {Mao} S.,    {White} S.~D.~M.,  1998, \mnras, 295, 319

\bibitem[\protect\citeauthoryear{{Moody}, {Guo}, {Mandelker}, {Ceverino},
  {Mozena}, {Koo}, {Dekel} \& {Primack}}{{Moody} et~al.}{2014}]{Moody}
{Moody} C.~E.,  {Guo} Y.,  {Mandelker} N.,  {Ceverino} D.,  {Mozena} M.,  {Koo}
  D.~C.,  {Dekel} A.,    {Primack} J.,  2014, \mnras, 444, 1389

\bibitem[\protect\citeauthoryear{{Morishita}, {Ichikawa} \& {Kajisawa}}{{Morishita}
  et~al.}{2014}]{Morishita14}  
{Morishita}, T., 
{Ichikawa}, T., \& {Kajisawa}, M.\ 2014, \apj, 785, 18  

\bibitem[\protect\citeauthoryear{{Moster}, {Naab} \& {White}}{{Moster}
  et~al.}{2013}]{Moster12}
{Moster} B.~P.,  {Naab} T.,    {White} S.~D.~M.,  2013, \mnras, 428, 3121

\bibitem[\protect\citeauthoryear{{Naab}, {Johansson}, {Ostriker} \&
  {Efstathiou}}{{Naab} et~al.}{2007}]{Naab07}
{Naab} T.,  {Johansson} P.~H.,  {Ostriker} J.~P.,    {Efstathiou} G.,  2007,
  \apj, 658, 710

\bibitem[\protect\citeauthoryear{{Neistein} \& {Dekel}}{{Neistein} \&
  {Dekel}}{2008}]{Neistein08}
{Neistein} E.,  {Dekel} A.,  2008, \mnras, 388, 1792

\bibitem[\protect\citeauthoryear{{Noguchi}}{{Noguchi}}{1999}]{Noguchi99}
{Noguchi} M.,  1999, \apj, 514, 77

\bibitem[\protect\citeauthoryear{{Ocvirk}, {Pichon} \& {Teyssier}}{{Ocvirk}
  et~al.}{2008}]{ocvirk08}
{Ocvirk} P.,  {Pichon} C.,    {Teyssier} R.,  2008, \mnras, 390, 1326

\bibitem[\protect\citeauthoryear{{Patel}, {van Dokkum}, {Franx}, {Quadri},
  {Muzzin}, {Marchesini}, {Williams}, {Holden} \& {Stefanon}}{{Patel}
  et~al.}{2013}]{Patel13}
{Patel} S.~G.,  {van Dokkum} P.~G.,  {Franx} M.,  {Quadri} R.~F.,  {Muzzin} A.,
   {Marchesini} D.,  {Williams} R.~J.,  {Holden} B.~P.,    {Stefanon} M.,
  2013, \apj, 766, 15

\bibitem[\protect\citeauthoryear{{Reddy}, {Steidel}, {Pettini}, {Adelberger},
  {Shapley}, {Erb} \& {Dickinson}}{{Reddy} et~al.}{2008}]{Reddy08}
{Reddy} N.~A.,  {Steidel} C.~C.,  {Pettini} M.,  {Adelberger} K.~L.,  {Shapley}
  A.~E.,  {Erb} D.~K.,    {Dickinson} M.,  2008, \apjs, 175, 48

\bibitem[\protect\citeauthoryear{{Robertson}, {Cox}, {Hernquist}, {Franx},
  {Hopkins}, {Martini} \& {Springel}}{{Robertson} et~al.}{2006}]{Robertson06}
{Robertson} B.,  {Cox} T.~J.,  {Hernquist} L.,  {Franx} M.,  {Hopkins} P.~F.,
  {Martini} P.,    {Springel} V.,  2006, \apj, 641, 21

\bibitem[\protect\citeauthoryear{{Scannapieco}, {White}, {Springel} \&
  {Tissera}}{{Scannapieco} et~al.}{2009}]{Scannapieco09}
{Scannapieco} C.,  {White} S.~D.~M.,  {Springel} V.,    {Tissera} P.~B.,  2009,
  \mnras, 396, 696

\bibitem[\protect\citeauthoryear{{Sersic}}{{Sersic}}{1968}]{Sersic}
{Sersic} J.~L.,  1968, {Atlas de galaxias australes}

\bibitem[\protect\citeauthoryear{{Shapiro}, {Genzel}, {F{\"o}rster Schreiber},
  {Tacconi}, {Bouch{\'e}}, {Cresci}, {Davies} \& {et al.,}}{{Shapiro}
  et~al.}{2008}]{Shapiro08}
{Shapiro} K.~L.,  {Genzel} R.,  {F{\"o}rster Schreiber} N.~M.,  {Tacconi}
  L.~J.,  {Bouch{\'e}} N.,  {Cresci} G.,  {Davies} R.,    {et al.,} 2008, \apj,
  682, 231

\bibitem[\protect\citeauthoryear{{Stark}, {Swinbank}, {Ellis}, {Dye}, {Smail}
  \& {Richard}}{{Stark} et~al.}{2008}]{Stark08}
{Stark} D.~P.,  {Swinbank} A.~M.,  {Ellis} R.~S.,  {Dye} S.,  {Smail} I.~R.,
  {Richard} J.,  2008, \nat, 455, 775

\bibitem[\protect\citeauthoryear{{Stewart}, {Bullock}, {Wechsler} \&
  {Maller}}{{Stewart} et~al.}{2009}]{Stewart09}
{Stewart} K.~R.,  {Bullock} J.~S.,  {Wechsler} R.~H.,    {Maller} A.~H.,  2009,
  \apj, 702, 307

\bibitem[\protect\citeauthoryear{{Szomoru}, {Franx}, {Bouwens}, {van Dokkum},
  {Labb{\'e}}, {Illingworth} \& {Trenti}}{{Szomoru} et~al.}{2011}]{Szomoru11}
{Szomoru} D.,  {Franx} M.,  {Bouwens} R.~J.,  {van Dokkum} P.~G.,  {Labb{\'e}}
  I.,  {Illingworth} G.~D.,    {Trenti} M.,  2011, \apjl, 735, L22

\bibitem[\protect\citeauthoryear{{Szomoru}, {Franx} \& {van Dokkum}}{{Szomoru}
  et~al.}{2012}]{Szomoru12}
{Szomoru} D.,  {Franx} M.,    {van Dokkum} P.~G.,  2012, \apj, 749, 121

\bibitem[\protect\citeauthoryear{{Toft}, {van Dokkum}, {Franx}, {Labbe},
  {F{\"o}rster Schreiber}, {Wuyts}, {Webb}, {Rudnick}, {Zirm}, {Kriek}, {van
  der Werf}, {Blakeslee}, {Illingworth}, {Rix}, {Papovich} \&
  {Moorwood}}{{Toft} et~al.}{2007}]{Toft07}
{Toft} S.,  {van Dokkum} P.,  {Franx} M.,  {Labbe} I.,  {F{\"o}rster Schreiber}
  N.~M.,  {Wuyts} S.,  {Webb} T.,  {Rudnick} G.,  {Zirm} A.,  {Kriek} M.,  {van
  der Werf} P.,  {Blakeslee} J.~P.,  {Illingworth} G.,  {Rix} H.-W.,
  {Papovich} C.,    {Moorwood} A.,  2007, \apj, 671, 285

\bibitem[\protect\citeauthoryear{{Toomre}}{{Toomre}}{1964}]{toomre64}
{Toomre} A.,  1964, \apj, 139, 1217

\bibitem[\protect\citeauthoryear{{Trujillo}, {F{\"o}rster Schreiber},
  {Rudnick}, {Barden}, {Franx}, {Rix}, {Caldwell} \& {et al.,}}{{Trujillo}
  et~al.}{2006}]{Trujillo06}
{Trujillo} I.,  {F{\"o}rster Schreiber} N.~M.,  {Rudnick} G.,  {Barden} M.,
  {Franx} M.,  {Rix} H.-W.,  {Caldwell} J.~A.~R.,    {et al.,} 2006, \apj, 650,
  18

\bibitem[\protect\citeauthoryear{{Tweed}, {Devriendt}, {Blaizot}, {Colombi} \&
  {Slyz}}{{Tweed} et~al.}{2009}]{Tweed09}
{Tweed} D.,  {Devriendt} J.,  {Blaizot} J.,  {Colombi} S.,    {Slyz} A.,  2009,
  \aap, 506, 647

\bibitem[\protect\citeauthoryear{{van den Bergh}}{{van den
  Bergh}}{1996}]{Bergh96}
{van den Bergh} S.,  1996, \aj, 112, 2634

\bibitem[\protect\citeauthoryear{{van Dokkum}, {Bezanson}, {van der Wel}
  \& {et al.,}}{{van Dokkum} et~al.}{2014}]{vDokkum14}
{van Dokkum}, P.~G., {Bezanson}, R., {van der Wel}, A., et al.\ 2014, \apj, 791, 45

\bibitem[\protect\citeauthoryear{{van Dokkum}, {Brammer}, {Fumagalli}, {Nelson}
  \& {et al.,}}{{van Dokkum} et~al.}{2011}]{vDokkum11}
{van Dokkum} P.~G.,  {Brammer} G.,  {Fumagalli} M.,  {Nelson} E.,    {et al.,}
  2011, \apjl, 743, L15

\bibitem[\protect\citeauthoryear{{van Dokkum}, {Franx}, {Kriek}, {Holden},
  {Illingworth}, {Magee}, {Bouwens}, {Marchesini}, {Quadri}, {Rudnick},
  {Taylor} \& {Toft}}{{van Dokkum} et~al.}{2008}]{vDokkum08}
{van Dokkum} P.~G.,  {Franx} M.,  {Kriek} M.,  {Holden} B.,  {Illingworth}
  G.~D.,  {Magee} D.,  {Bouwens} R.,  {Marchesini} D.,  {Quadri} R.,  {Rudnick}
  G.,  {Taylor} E.~N.,    {Toft} S.,  2008, \apjl, 677, L5

\bibitem[\protect\citeauthoryear{{van Dokkum}, {Whitaker}, {Brammer}, {Franx},
  {Kriek}, {Labb{\'e}}, {Marchesini}, {Quadri}, {Bezanson}, {Illingworth},
  {Muzzin}, {Rudnick}, {Tal} \& {Wake}}{{van Dokkum} et~al.}{2010}]{vDokkum10}
{van Dokkum} P.~G.,  {Whitaker} K.~E.,  {Brammer} G.,  {Franx} M.,  {Kriek} M.,
   {Labb{\'e}} I.,  {Marchesini} D.,  {Quadri} R.,  {Bezanson} R.,
  {Illingworth} G.~D.,  {Muzzin} A.,  {Rudnick} G.,  {Tal} T.,    {Wake} D.,
  2010, \apj, 709, 1018

\bibitem[\protect\citeauthoryear{{Weinzirl}, {Jogee}, {Conselice}, {Papovich},
  {Chary}, {Bluck}, {Gr{\"u}tzbauch}, {Buitrago}, {Mobasher}, {Lucas},
  {Dickinson} \& {Bauer}}{{Weinzirl} et~al.}{2011}]{Weinzirl11}
{Weinzirl} T.,  {Jogee} S.,  {Conselice} C.~J.,  {Papovich} C.,  {Chary} R.-R.,
   {Bluck} A.,  {Gr{\"u}tzbauch} R.,  {Buitrago} F.,  {Mobasher} B.,  {Lucas}
  R.~A.,  {Dickinson} M.,    {Bauer} A.~E.,  2011, \apj, 743, 87

\bibitem[\protect\citeauthoryear{{Williams}, {Giavalisco}, {Cassata}, {Tundo},
  {Wiklind} \& {et al.,}}{{Williams} et~al.}{2014}]{Williams14}
{Williams} C.~C.,  {Giavalisco} M.,  {Cassata} P.,  {Tundo} E.,  {Wiklind} T.,
    {et al.,} 2014, \apj, 780, 1

\bibitem[\protect\citeauthoryear{{Williams}, {Quadri}, {Franx}, {van Dokkum},
  {Toft}, {Kriek} \& {Labb{\'e}}}{{Williams} et~al.}{2010}]{Williams10}
{Williams} R.~J.,  {Quadri} R.~F.,  {Franx} M.,  {van Dokkum} P.,  {Toft} S.,
  {Kriek} M.,    {Labb{\'e}} I.,  2010, \apj, 713, 738

\bibitem[\protect\citeauthoryear{{Wuyts}, {Cox}, {Hayward}, {Franx},
  {Hernquist}, {Hopkins}, {Jonsson} \& {van Dokkum}}{{Wuyts}
  et~al.}{2010}]{Wuyts10}
{Wuyts} S.,  {Cox} T.~J.,  {Hayward} C.~C.,  {Franx} M.,  {Hernquist} L.,
  {Hopkins} P.~F.,  {Jonsson} P.,    {van Dokkum} P.~G.,  2010, \apj, 722, 1666

\bibitem[\protect\citeauthoryear{{Wuyts}, {F{\"o}rster Schreiber}, {Genzel},
  {Guo}, {Barro} \& {et al.,}}{{Wuyts} et~al.}{2012}]{Wuyts12}
{Wuyts} S.,  {F{\"o}rster Schreiber} N.~M.,  {Genzel} R.,  {Guo} Y.,  {Barro}
  G.,    {et al.,} 2012, \apj, 753, 114


\end{thebibliography}

\bsp

\label{lastpage}

\end{document}